\documentclass[notitlepage,prd,amsfonts,noshowpacs,nofootinbib]{revtex4-1}
\usepackage{amsmath,graphicx}
\usepackage {hyperref}

\begin{document}

\title{Non-Gaussianities in DBI inflation with angular motion}

\author{Taichi Kidani\footnote{Taichi.Kidani@port.ac.uk}}

\author{Kazuya Koyama\footnote{Kazuya.Koyama@port.ac.uk}}

\affiliation{Institute of Cosmology and Gravitation, University of Portsmouth, Portsmouth PO1 3FX, UK.}

\date{\today}
\begin{abstract}
We study DBI spinflation models with angular potentials that are derived in string theory. We analyse the background dynamics with different parameter sets and study the impact of changing each parameter on inflationary dynamics. It is known that the conversion of the entropy perturbation into the curvature perturbation gives multi-field DBI inflation models a possibility of satisfying the observational constraints by relaxing the stringent microphysical constraint that disfavours single field DBI inflation models. We show that our model is excluded by the Planck satellite observations even with the conversion mechanism regardless of the parameter set. 
\end{abstract}

\maketitle

\section{Introduction}
In the Inflationary scenario, the Cosmic Microwave Background (CMB) anisotropies are generated as a result of the quantum fluctuations of an inflaton field which drives inflation. The recent observations of the nearly Gaussian CMB anisotropies support the inflationary scenario because inflation naturally produces nearly Gaussian quantum fluctuations \cite{Komatsu:2010}. The statistical properties of the CMB anisotropies can be used to rule out some inflation models and a variety of inflation models have been studied using the results of the CMB observations \cite{Naruko:2009, Kaiser:2013, Kaiser:2014, Schutz:2014}. Among the observables that quantify the statistical properties of the CMB anisotropies, the non-Gaussianities has been studied extensively in the recent literature. The values of the non-Gaussianity parameters have been constrained by the Planck satellite observations to the unprecedented precision \cite{Ade:2013}. 

In many inflation models, the origin of inflaton is not specified. The Dirac-Born-Infeld (DBI) inflation \cite{Silverstein:2004, Alishahiha:2004}  motivated by string theory is driven by inflaton that is one of the scalar fields describing the position of a D-brane in the higher dimensional space in the effective four-dimensional theory. Also, DBI inflation has interesting features such as the speed limiting effect on the velocity of the scalar fields \cite{Underwood:2008} and large non-Gaussianity \cite{Chen:2006nt}. The single field Ultra-Violet (UV) DBI inflation models where a D3 brane is moving down the warped throat have been ruled out because of the microphysical constraint on the variation of inflaton in a string theory set-up combined with the constraints on the tensor-to-scalar ratio, the spectral index and non-Gaussianity by CMB observations \cite{Baumann:2007, Lidsey:2007, Ian:2008, Kobayashi:2007hm, Bean:2008}. 

However, DBI inflation naturally has more than one field because it is motivated by string theory in which there are six extra dimensions, which are the radial direction and five angular directions in the internal space. In the multi-field DBI inflation, the stringent constraints introduced above are relaxed as follows. A turn in the trajectory in the field space converts the entropy perturbation into the curvature perturbation on super-horizon scales \cite{Gordon:2001}. When we have such a conversion, it was shown that the constraints on single field DBI models are relaxed \cite{Langlois:2008wt, Langlois:2008qf, Langlois:2009, Arroja:2008yy}. However, we need a concrete model to make definite predictions on the values of the CMB observables. In \cite{Kidani:2012}, a toy model of two-field DBI inflation with a waterfall phase transition was studied. It was shown that this two-field DBI model satisfies the microphysical constraints combined with the WMAP observations even though it was ruled out with the PLANCK satellite observation. Because DBI inflation models are motivated by string theory, it is important to study models with potentials that are derived in string theory. 

In \cite{Baumann:2007ah}, the potential with the angular directions for a D3 brane in the warped deformed conifold was calculated. The impact of angular motion on DBI inflation has been studied with numerical calculations in \cite{spinflationbefore, spinflation}. In \cite{spinflation}, concrete angular dependent potentials in the spinflation models are derived and the background dynamics has been studied numerically. The rotational motion due to the angular potential in those models can potentially relax the microphysical constraints on the DBI inflation and make the values of the CMB observables compatible with the observations. We need numerical calculations to see if there is a parameter set which makes this model compatible with the PLANCK satellite observations. 

In this paper, the two-field DBI inflation model with the spinflation potential is analyzed. We perform numerical calculations in order to obtain the predictions for the CMB observables such as the power spectrum of the curvature perturbation, spectral index, tensor-to-scalar ratio and non-Gaussianity to see if this model is compatible with the microphysical constraint on DBI inflation and the CMB observations. In DBI inflation models, large equilateral type non-Gaussianities are generated from the bispectrum of the quantum fluctuations of the scalar fields in sub-horizon scales \cite{Alishahiha:2004}. On the other hand, the local type non-Gaussianities \cite{local} can be generated on super-horizon scales when there is a conversion from the entropy perturbation to the curvature perturbation (see \cite{Wands:2010af} for a review and references therein). Therefore, multi-field DBI inflation models are generally capable of producing a combination of the non-negligible equilateral type and local type non-Gausianities. With this feature, we can distinguish DBI models from other inflationary models \cite{Koyama:2010, Babich:2004, Creminelli:2006}. 

Both equilateral and local type non-Gaussianities in multi-field DBI inflation were first calculated in \cite{RenauxPetel:2009sj} assuming only an effective single field potential until the end of inflation where a tachyonic instability occurs in the angular direction. In this paper, we consider the dynamics in the angular direction during inflation because the potential has the angular dependence derived in string theory. 

In section \ref{sec:themodel}, we study the background dynamics of the spinflation model and show the numerical results of the background inflationary trajectories with different parameter sets after reviewing the background dynamics of the DBI inflation. In section \ref{sec:linearperturbation}, we review the linear perturbation theory of inflation and the conversion of the entropy perturbation into the curvature perturbation due to a turn of the background trajectory in the field space. Then, we show the analytic formulae of the spinflation that are valid when the dynamics is effectively single field around horizon crossing. We confirm those formulae by solving the linear perturbations numerically. In section \ref{sec:nongaussianity}, after reviewing the equilateral non-Gaussianity, we introduce the microphysical constraint that strongly disfavours the single field DBI inflation. Finally, we show that the spinflation model studied in this paper is excluded by the PLANCK satellite observations using the constraint on the CMB observables including the equilateral non-Gaussianity even with the conversion of the entropy perturbation into the curvature perturbation. In the last section, we summarise our work and discuss the cases in which we cannot use the analytic formulae. 

\section{Background dynamics}
\label{sec:themodel}
In this section, we first introduce multi-field DBI inflation. We briefly review a simple model of single field DBI inflation analysed in \cite{Silverstein:2004, Alishahiha:2004} because the two-field model that we study in this paper is approximated with this model in the region with a large radial coordinate when the dynamics is mainly in the radial direction. After introducing the spinflation model derived in \cite{spinflation}, we show our numerical results for the background trajectories with different values of the parameters. 

\subsection{Multi-field DBI inflation}
We consider the bulk whose warped geometry is given by \cite{Langlois:2008qf}
\begin{equation}\label{dbigeneralmetric}
ds^{2} = h^{-1/2} \left(y^{K}\right) g_{\mu \nu} dx^{\mu} dx^{\nu} + h^{1/2} \left(y^{K} \right) G_{IJ} \left(y^{K} \right) dy^{I} dy^{J} \equiv H_{AB} dY^{A} dY^{B}, 
\end{equation}
where $Y^{A}=\left\{x^{\mu}, y^{I} \right\}$ with the indices $\mu=0,1,2,3$ and $I=1,...,6$. $Y_{(b)}^{A}\left(x^{\mu}\right) = \left(x^{\mu},  \eta^{I}\left(x^{\mu}\right) \right)$ are the ten dimensional coordinates which specify the brane position in the bulk where $x^{\mu}$ is the four dimensional space-time coordinates on the brane. Then, the Lagrangian for the multi-field DBI inflation is given by
\begin{equation}
 P(X^{IJ},\phi^{I}) = \tilde{P}(\tilde{X},\phi^{I}) = - \frac{1}{f(\phi^{I})} \left(\sqrt{1-2 f(\phi^{I}) \tilde{X}} - 1\right) - V\left(\phi^{I}\right),
\label{multifieldaction}
\end{equation}
where $\phi^{I}$ are the scalar fields $(I=1,2,...)$ defined as
\begin{equation}\label{scalarfieldwithbranetension}
\phi^{I} = \sqrt{T_{3}} \eta^{I},
\end{equation}
where $T_3$ is the brane tension. Note that $\tilde{X}$ is defined in terms of the determinant \begin{eqnarray}
\mathcal{D} &=& \mbox{det} (\delta^{I}_{J} - 2 f X^{I}_{J} )\nonumber \\
&=& 1 - 2 f G_{IJ} X^{IJ} + 4 f^{2} X^{[I}_{I} X^{J]}_{J} - 8 f^{3} X^{[I}_{I} X^{J}_{J} X^{K]}_{K} + 16 f^{4} X^{[I}_{I} X^{J}_{J} X^{K}_{K} X^{L]}_{L}\, ,
\label{determinant}
\end{eqnarray}
as
\begin{equation}
 \tilde{X} = \frac{(1-\mathcal{D})}{2 f},
\end{equation}
where
\begin{equation}
 X^{IJ} \equiv - \frac{1}{2} \partial _{\mu} \phi^{I} \partial ^{\mu} \phi^{J}.
\end{equation}
Note that $f(\phi^{I})$ is defined
by the warp factor  $h(\phi^{I})$ and the brane tension $T_3$ as
\begin{equation}\label{warpfactorandbranetension}
 f(\phi^{I}) \equiv \frac{h(\phi^{I})}{T_{3}}.
\end{equation}
The sound speed is defined as
\begin{equation}
 c_{s} \equiv \sqrt{\frac{\tilde{P}_{,\tilde{X}}}{\tilde{P}_{,\tilde{X}} + 2 \tilde{X} \tilde{P}_{,\tilde{X} \tilde{X}}}} = \sqrt{1- 2 f \tilde{X}},
\end{equation}
where $,_{\tilde{X}}$ means the partial derivative with respect to $\tilde{X}$. Note that $\tilde{X}$ coincides with $X \equiv G_{IJ} X^{IJ}$ in the homogeneous background because all the spatial derivatives vanish. From the action (\ref{multifieldaction}),
we can show that
\begin{equation}
 \tilde{P}_{,\tilde{X}} = \frac{1}{c_{s}}.
\end{equation}
In this paper, we consider the Einstein-Hilbert action for gravity and hence all equations of motion are derived from the action
\begin{equation}
 S = \frac{1}{2} \int d^{4}x \sqrt{-g} \left[^{(4)}R + 2 P(X^{IJ},\phi^{I}) \right],
\label{actionofdbi}
\end{equation}
where we set $8 \pi G = 1$ and $^{(4)}R$ is the four dimensional Ricci curvature. The Friedmann equation is given by
\begin{equation}
 3 M^{2}_{P} H^{2} = \frac{1}{f(\phi)} \left(\frac{1}{c_{s}} - 1 \right) + V(\phi^{I}).\label{friedmann}
\end{equation}
where $H$ is the Hubble parameter $H \equiv \dot{a}/a$ with the scale factor $a$. 

\subsection{Single field DBI inflation with a quadratic potential}
\label{singlefieldquadratic}
Here, we introduce the single field DBI inflation model in a simple set-up in string theory following \cite{Silverstein:2004, Alishahiha:2004}. In the AdS throat, a potential of the mass term is given by
\begin{equation}\label{dbisingleexamplepotential}
V \left(\phi \right) = m^{2} \phi^{2},
\end{equation}
with the mass of inflaton $m$. For a pure AdS$_{5}$ throat of radius $R$, we have
\begin{equation}\label{dbisingleexamplewarp}
f \left(\phi \right) = \frac{\lambda}{\phi^{4}},
\end{equation}
where $\lambda \equiv R^{4}/\alpha'^{2}$ is constant with the inverse string tension $\alpha'$. Note that the throats arising from IIB flux compactifications can look approximately AdS  \cite{Alishahiha:2004}. Let us define the slow-roll parameters as
\begin{equation}
 \epsilon = - \frac{\dot{H}}{H^{2}}, \:\:\:\:\: \eta = \frac{\dot{\epsilon}}{H \epsilon}, \:\:\:\:\: s = \frac{\dot{c_{s}}}{H c_{s}}.
\label{slowrollparameters}
\end{equation}
Then, when the field value is sub-Planckian, namely $\phi / M_{P} \ll 1$, it is shown that we have the relations \cite{Silverstein:2004, Alishahiha:2004}
\begin{equation}\label{alltheresultsindbisingleexample}
\phi = \frac{\sqrt{\lambda}}{t}, \:\:\:\:\: \gamma = c^{-1}_{s} = \sqrt{\frac{4}{3 \lambda}} M_{P} m t^{2}, \:\:\:\:\: H = \frac{1}{\epsilon t}, \:\:\:\:\: \epsilon = \sqrt{\frac{3}{\lambda}}\frac{M_{P}}{m},
\end{equation}
if we assume that the sound speed $c_{s}$ is much smaller than unity where $M_{P}$ is the Planck mass. Then, the slow roll parameters $\eta$ and $s$ are given by
\begin{equation}\label{relationchapthreesimple}
 \eta = 0, \:\:\:\:\: s = -2 \epsilon,
\end{equation}
from Eqs. (\ref{slowrollparameters}) and (\ref{alltheresultsindbisingleexample}). Therefore, in this model, the background dynamics can be expressed with these functions of time when the field value is sub-Planckian and the sound speed is much smaller than unity. 

\subsection{Spinflation model}
\label{twofield}
Let us introduce the two-field model with a potential derived in string theory embedding a warped throat into a compact Calabi-Yau space with all moduli fields stabilized following \cite{spinflation}. Let us define $\chi$ and $\theta$ to be the radial and angular coordinates in the warped throat in the internal space respectively. The field space metric is then given by
\begin{equation}
ds^{2}=\tilde{g}_{mn}dy^{m}dy^{n}=\kappa^{4/3}\left[\frac{d\chi^{2}}{6K\left(\chi\right)^{2}}+B\left(\chi\right)d\theta^{2}\right],
\end{equation}
with
\begin{equation}
K\left(\chi\right)=\frac{\left(\sinh{\chi}\cosh{\chi}-\chi\right)^{1/3}}{\sinh{\chi}},\,\,\,\,\,B\left(\chi\right)=\frac{1}{2}K\left(\chi\right)\cosh{\chi}, 
\end{equation}
and the deformation parameter $\kappa$. Note that $\eta^{I} = \left(\chi, \theta\right)$ in Eq. (\ref{scalarfieldwithbranetension}). The warp factor in this model is given by \cite{Klebanov:2000}
\begin{equation}\label{warpfactor}
h\left(\chi\right) \equiv e^{-4A} = 2\left(g_{\rm{s}}M\alpha '\right)^{2}\kappa^{-8/3}I\left(\chi\right),
\end{equation}
where
\begin{equation}\label{eqforfunctioni}
I\left(\chi\right)\equiv\int^{\infty}_{\chi}dx\frac{x\coth{x}-1}{\sinh^2{x}}\left(\sinh{x}\cosh{x}-x\right)^{1/3},
\end{equation}
with the parameter $M$, the string coupling $g_{\rm{s}}$ and the inverse string tension $\alpha'$. By varying the action (\ref{actionofdbi}) with respect to the fields, the equations of motion for the radial scalar field $\chi$ and angular scalar field $\theta$ are given by 
\begin{equation}\label{equationofmotionchibackground}
\begin{split}
\ddot{\chi}=&-\frac{3H}{\gamma^{2}}\dot{\chi}-4A_{,\chi}\left(\gamma^{-1}-1\right)\dot{\chi}^{2}-12\kappa^{-4/3}K^{2}A_{,\chi}e^{4A}\left(\gamma^{-1}-1\right)^{2}\\
&+\frac{K_{,\chi}}{K}\dot{\chi}^{2}+3K^{2}B_{,\chi}\dot{\theta}^{2}+e^{-4A}\dot{\theta}\dot{\chi}\frac{U_{,\theta}}{\gamma}-\left(6K^{2}\kappa^{-4/3}-e^{-4A}\dot{\chi}^{2}\right)\frac{U_{,\chi}}{\gamma},
\end{split}
\end{equation}
\begin{equation}\label{equationofmotionthetabackground}
\ddot{\theta}=-\frac{3H}{\gamma^{2}}\dot{\theta}-4A_{,\chi}\left(\gamma^{-1}-1\right)\dot{\chi}\dot{\theta}-\frac{B_{,\chi}}{B}\dot{\chi}\dot{\theta}+e^{-4A}\dot{\chi}\dot{\theta}\frac{U_{,\chi}}{\gamma}-\left(\frac{\kappa^{-4/3}}{B}-e^{-4A}\dot{\theta}^{2}\right)\frac{U_{,\theta}}{\gamma},
\end{equation}
where the subscripts denote derivatives with respect to the fields as $_{,\chi}\equiv\frac{\partial}{\partial\chi}$ and $_{,\theta}\equiv\frac{\partial}{\partial\theta}$. The sound speed of the scalar field perturbations is given by
\begin{equation}\label{spinflationsoundspeedandfactor}
\begin{split}
c_{\rm{s}}=\gamma^{-1}&=\sqrt{1-e^{-4A}\tilde{g}_{mn}\dot{y}^{m}\dot{y}^{n}}\\
 & = \sqrt{1 - h \kappa^{4/3} \left(\frac{\dot{\chi}^{2}}{6K^{2}}+B \dot{\theta}^{2} \right)}, 
\end{split}
\end{equation}
where $y^{\rm{m}}=\left(\chi,\,\theta\right)$. 
The potential is derived by solving the equation of motion for $\Phi_{-}$
\begin{equation}\label{equationofmotionforpotentialdependingonembedding}
\nabla^{2} \Phi_{-} = \frac{g_{s}}{24 h^{2}} \left\lvert G_{-} \right\rvert + \mathcal{R}_{4} + h \left\lvert \nabla \Phi_{-} \right\rvert^{3} + \mathcal{S}_{\rm{local}},
\end{equation}
which is derived from the IIB supergravity action \cite{Baumann:2010} where $\nabla^{2}$ is the Laplacian with respect to the 6D metric $G_{IJ}$ in Eq. (\ref{dbigeneralmetric}), $h$ is the warp factor, $g_{s} = e^{\phi}$ is the string coupling with the dilation field $\phi$, $ \mathcal{R}_{4}$ is the four dimensional Ricci scalar, $\mathcal{S}_{\rm{local}}$ is the localised sources and 
\begin{equation}
G_{-} = \left(\star_{6} - i \right) G_{3},
\end{equation}
is the imaginary anti-self-dual (IASD) component of the complex three form flux $G_{3}$ where $\star_{6}$ is the six dimensional Hodge star operator. In \cite{spinflation}, the equation of motion (\ref{equationofmotionforpotentialdependingonembedding}) is solved including linearised perturbations around the ISD solution with a warped throat embedded into a compact Calabi-Yau space with all moduli fields stabilised. Because the dominant source for $\Phi_{-}$ is $G_{-}$ and it sources only second order perturbations, the perturbations of $\Phi_{-}$ around the ISD condition $\left(\Phi_{-}=0\right)$ satisfy
\begin{equation}\label{firsteqodspinlaplace}
\nabla^{2} \Phi_{-} = 0,
\end{equation}
at the linear level. For a general warped deformed conifold, the Laplacian of Eq. (\ref{firsteqodspinlaplace}) takes a simple form when we are only interested in the low lying states that are dependent on only one angular coordinate $\theta$. In this case, the leading order term of the eigenfunction with the lowest angular mode $\ell=1$ is given by
\begin{equation}\label{eigenfunctionspin}
\Phi_{-} \propto \left(\cosh{\chi} \sinh{\chi} - \chi \right)^{1/3} \cos{\theta}.
\end{equation}
Adding the mass term that arises from the effects of the bulk geometry, the potential is derived with the eigenfunction (\ref{eigenfunctionspin}) as
\begin{equation}\label{potential}
V\left(\phi^{I} \right) = T_{3} U = T_{3} \left[\frac{1}{2}m_{0}^{2}\left\{r\left(\chi\right)^{2}+c_{2}K\left(\chi\right)\sinh{\eta}\cos{\theta}\right\}+U_{0} \right],
\end{equation}
where the proper radial coordinate is given by
\begin{equation}\label{eqforproperradial}
r\left(\chi\right)=\frac{\kappa^{2/3}}{\sqrt{6}}\int^{\chi}_{0}\frac{dx}{K\left(x\right)},
\end{equation}
with an arbitrary constant $c_{2}$ which is of the same order as the deformation parameter: $c_{2} \sim \kappa^{4/3}$. Note that the constant $U_{0}$ is chosen so that the global minimum of $V$ is $V=0$ . 

\begin{figure}[!htb]
\centering
\includegraphics[width=12cm]{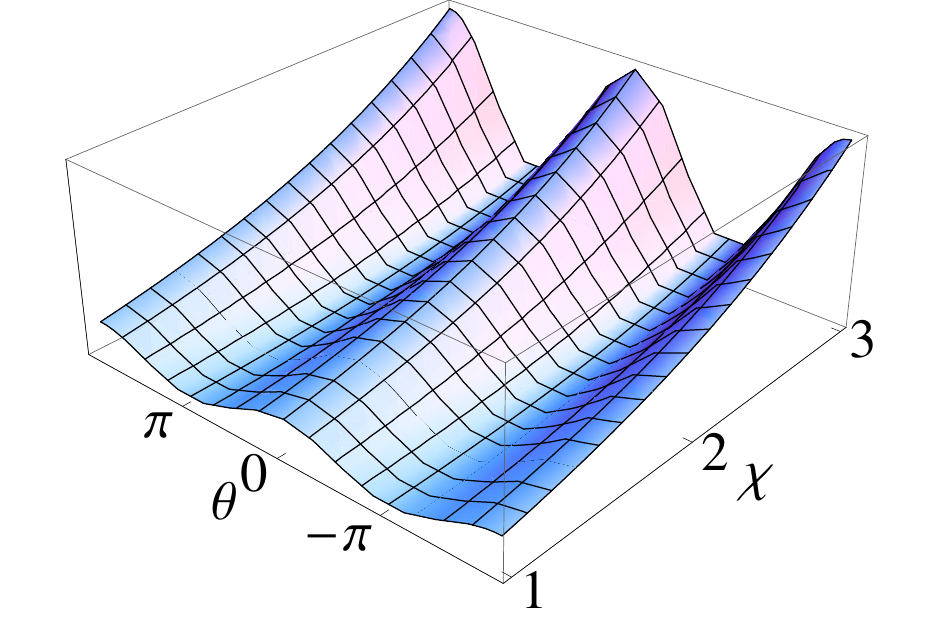}
\caption[Spinflation potential]{The potential (\ref{potential}) with $g_{\rm{s}}=1/2\pi\,\left[\mathcal{M}^{0} \right]$, $\alpha'=1\,\left[\mathcal{M}^{-2} \right]$, $M=10^{6}\pi\,\left[\mathcal{M}^{0} \right]$, $\kappa=10^{-11}\,\left[\mathcal{M}^{-3/2} \right]$ and $m_{0}=4.5\times10^{-5}\,\left[\mathcal{M} \right]$. Note $\mathcal{M}$ is the mass unit defined in Eq. (\ref{massunitdefinitionspin}). The minima of the potential in the angular direction are at $\theta=\left(2N+1\right)\pi$ with an integer $N$ while the maxima are along the lines $\theta = 2N \pi$. The radial potential is quadratic in terms of $r\left(\chi\right)$.}{\label{spinflationpotentialshape}}
\end{figure}

Figure \ref{spinflationpotentialshape} shows the shape of the potential. From Eqs. (\ref{scalarfieldwithbranetension}), (\ref{warpfactorandbranetension}), (\ref{friedmann}) and (\ref{potential}), the Friedmann equation reads
\begin{equation}\label{friedmannspinflation}
H^{2}=\frac{T_{3}}{3M^{2}_{P}}\left[\frac{1}{h}\left(\gamma-1\right)+U\right]. 
\end{equation}
Combining the time derivative of Eq. (\ref{friedmannspinflation}) with respect to the cosmic time $t$ and the continuity equation $\dot{\rho} = - 3 H \left(E + P \right)$, we obtain
\begin{equation}\label{spinflationhubblederivative}
\dot{H} = - \frac{T_{3}}{2 M^{2}_{P} h} \left(\gamma - \gamma^{-1} \right). 
\end{equation}
Note that $E$ denotes the energy density which is equivalent to $3M^{2}_{P} H^{2}$ while $P$ is the pressure which is equivalent to the Lagrangian given in Eq. (\ref{multifieldaction}). In \cite{Baumann:2007}, it is shown that the volume of the compactification is constrained by the Planck mass. Because the volume of the warped throat must be smaller than the total volume of the internal space, we have the relation
\begin{equation}\label{planckmassbound}
M^{2}_{P} \geq \frac{\kappa^{4/3}g_{\rm{s}}M^{2}T_{3}}{6 \pi} J\left(\chi_{\rm{UV}} \right),
\end{equation}
where $J\left(\chi \right) = \int d\chi I \left(\chi \right) \sinh^{2}{\chi}$ with the UV cut-off of the throat at $\chi = \chi_{\rm{UV}}$. Note that $T_{3}$ is the D3-brane tension given in Eq. (\ref{scalarfieldwithbranetension}). The mass dimension $\left[\mathcal{M} \right]$ is determined by the relation (\ref{planckmassbound}), which is rewritten as
\begin{equation}\label{massunitdefinitionspin}
M^{2}_{P} = \frac{\bar{\kappa}^{4/3}g_{\rm{s}}M^{2}\bar{T}_{3}}{6 \pi} N J\left(\chi_{\rm{UV}} \right) \mathcal{M}^{2},
\end{equation}
with the dimensionless parameters $N\geq1$, $\bar{T}_{3} = T_{3}/\mathcal{M}^{4}$ and $\bar{\kappa}^{4/3} = \kappa^{4/3} \mathcal{M}^{2}$. We use only the dimensionless parameters in the numerical calculations and that $g_{\rm{s}}$ and $M$ are dimensionless. Below, all the values of the parameters are given in the mass unit $\left[\mathcal{M} \right]$ unless stated otherwise. The dimensions of the parameters are: $g_{\rm{s}}\,\left[\mathcal{M}^{0} \right]$, $M\,\left[\mathcal{M}^{0} \right]$, $T_{s}\,\left[\mathcal{M}^{4} \right]$, $\kappa^{4/3}\,\left[\mathcal{M}^{-2} \right]$ and $\alpha'\,\left[\mathcal{M}^{-2} \right]$. By saturating the Planck mass bound (\ref{planckmassbound}), Eqs. (\ref{friedmannspinflation}) and (\ref{spinflationhubblederivative}) are rewritten as
\begin{equation}\label{spinflationsaturatedfriedmann}
H^{2}=\frac{T_{3}}{3M^{2}_{P}}\left[\frac{1}{h}\left(\gamma-1\right)+U\right] \rightarrow \frac{2 \pi}{\kappa^{4/3}g_{\rm{s}}M^{2}J_{\rm{UV}}} \left[\frac{1}{h}\left(\gamma-1\right)+U\right], 
\end{equation}
\begin{equation}\label{spinflationsaturatedhubblederivative}
\dot{H} = - \frac{T_{3}}{2 M^{2}_{P} h} \left(\gamma - \gamma^{-1} \right) \rightarrow - \frac{3 \pi}{\kappa^{4/3}g_{\rm{s}}M^{2}J_{\rm{UV}} h} \left(\gamma - \gamma^{-1} \right).
\end{equation}

\subsection{Numerical results for the background trajectories}\label{subsec:backgroundbehaviourspin}
\begin{figure}
  \centering
   \begin{tabular}{cc} 
    \includegraphics[width=70mm]{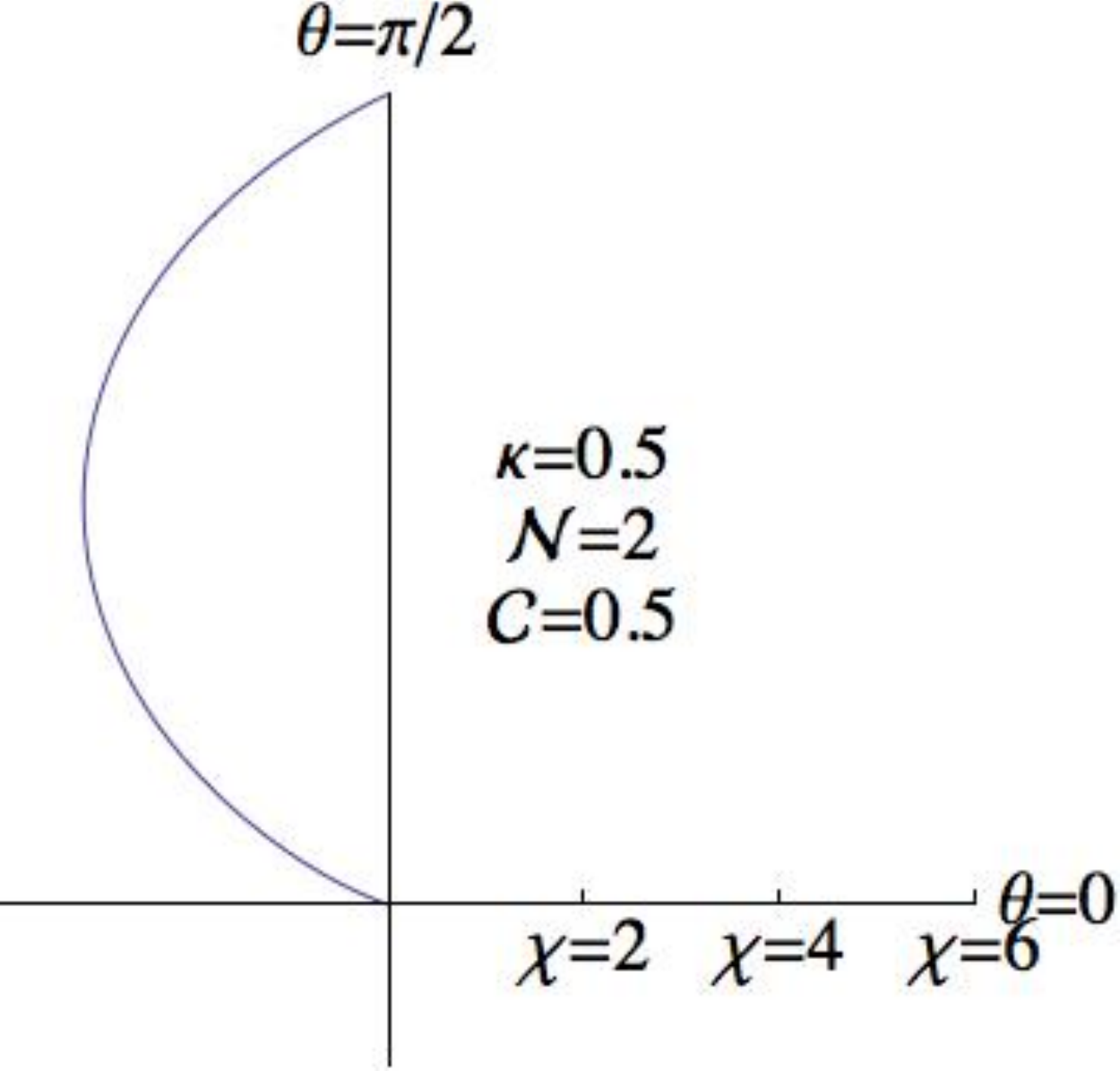}&
    \includegraphics[width=70mm]{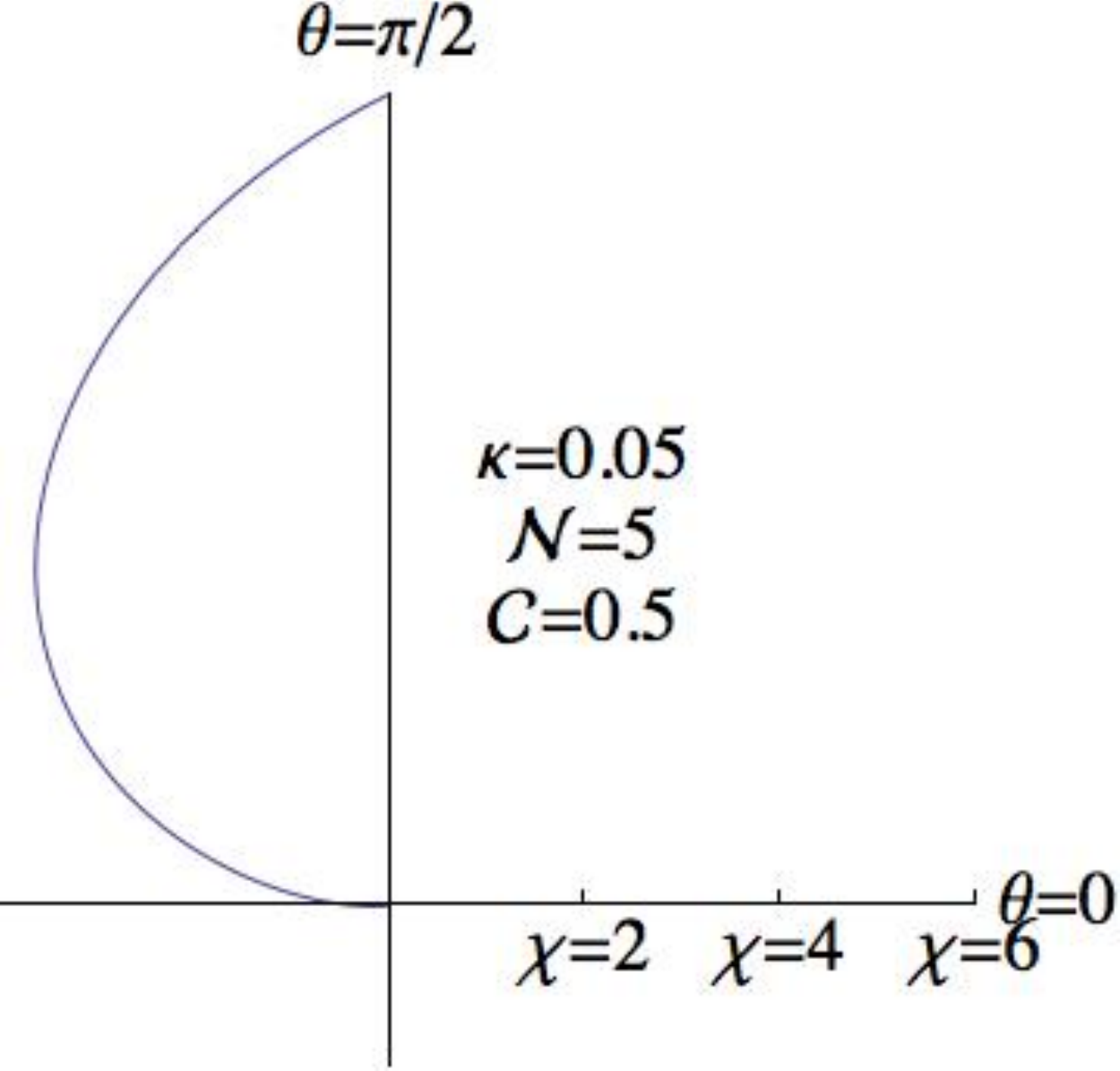}\\
    \includegraphics[width=70mm]{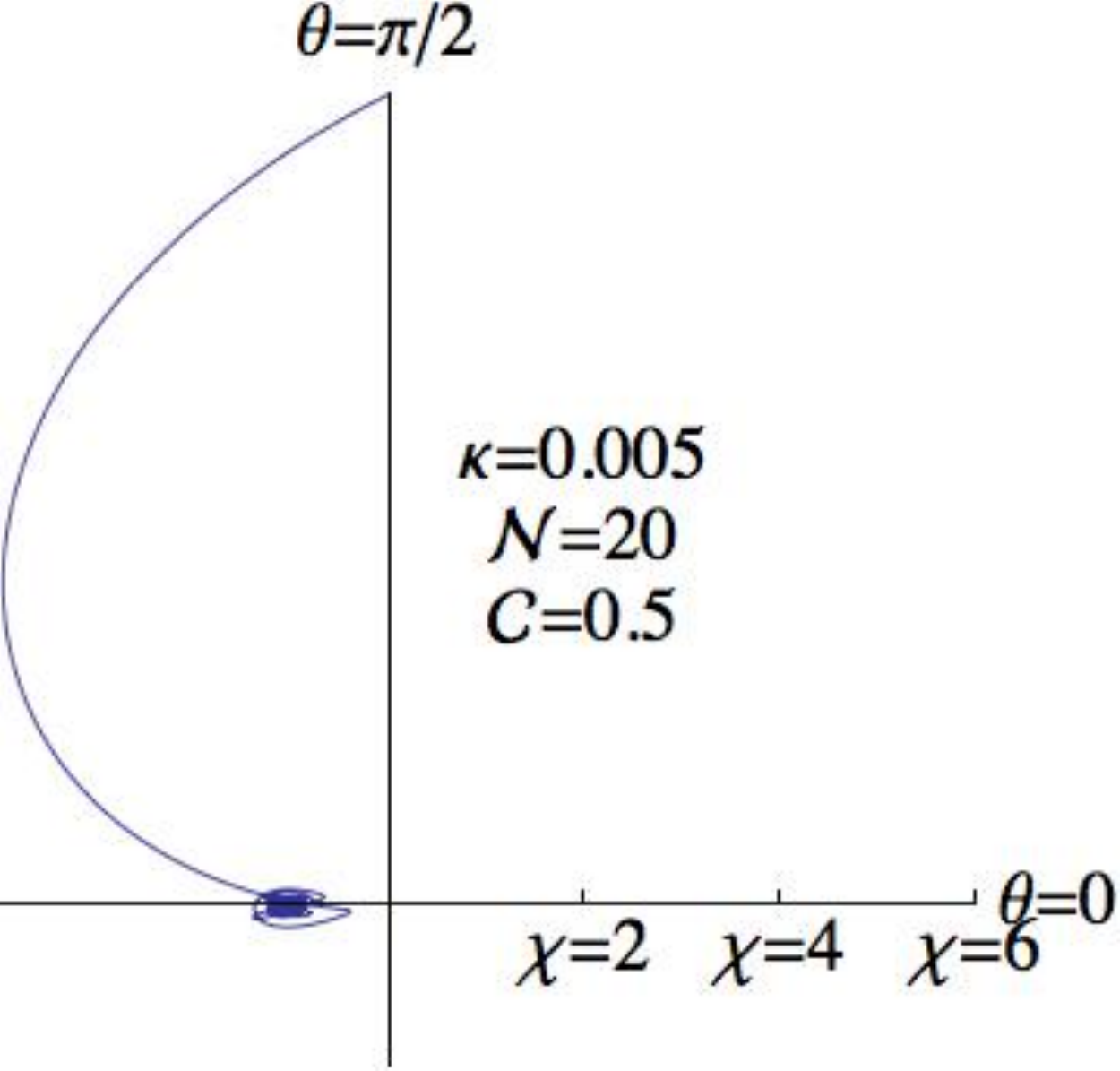}&
    \includegraphics[width=70mm]{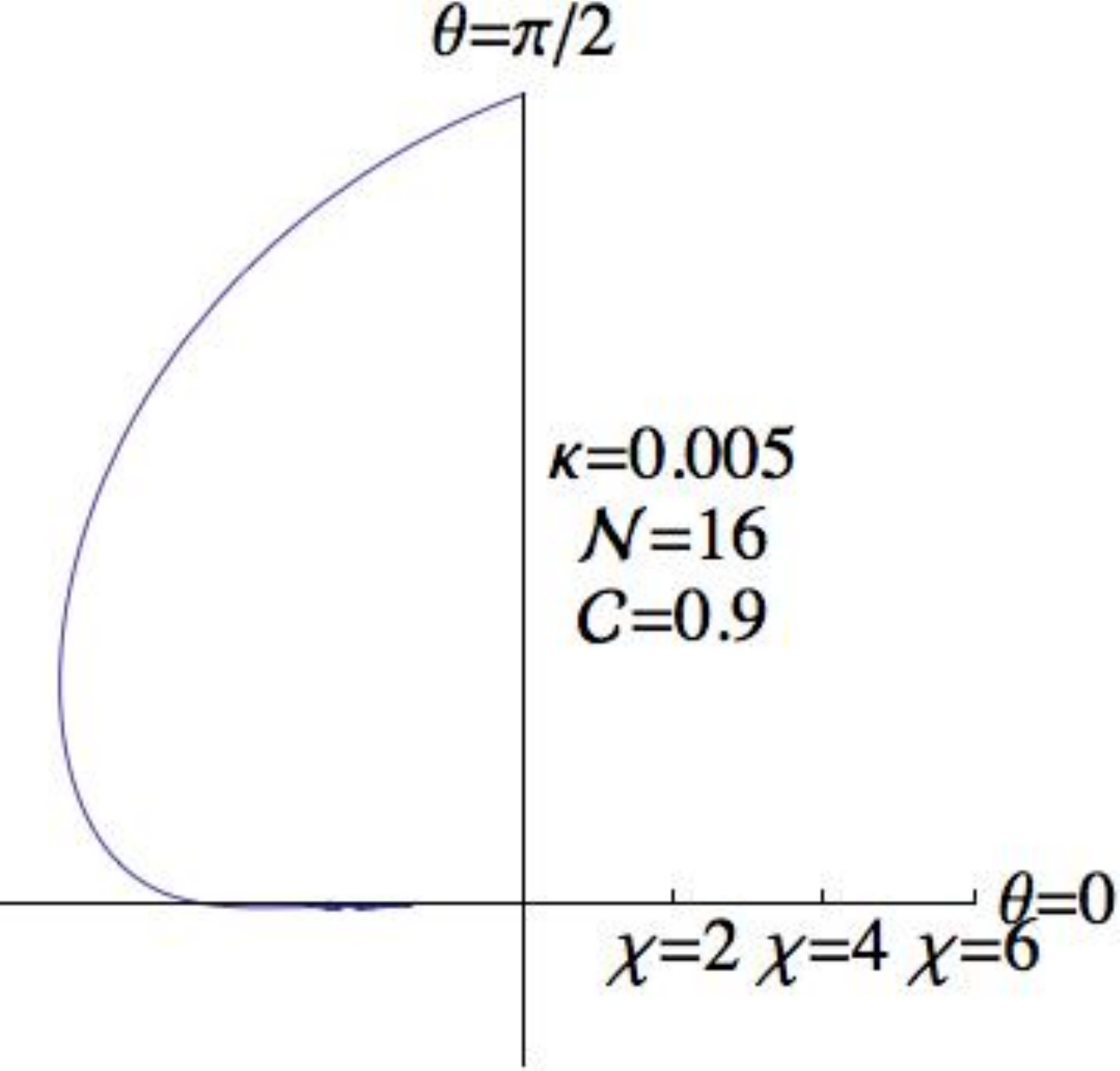}\\
    \includegraphics[width=70mm]{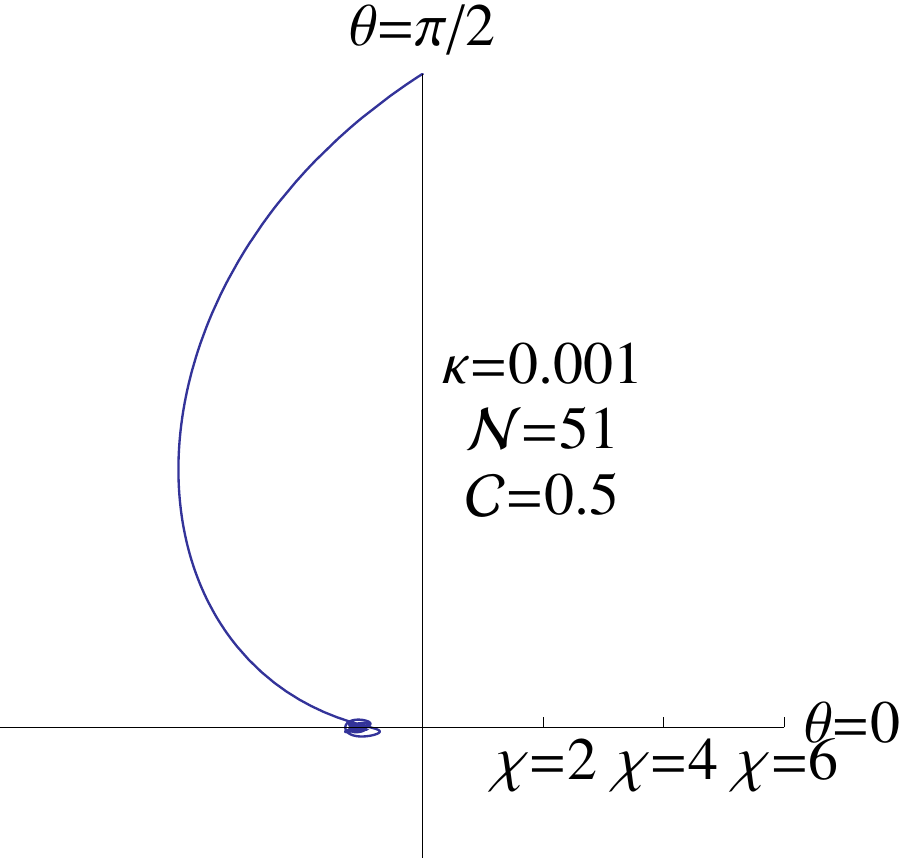}&
    \includegraphics[width=70mm]{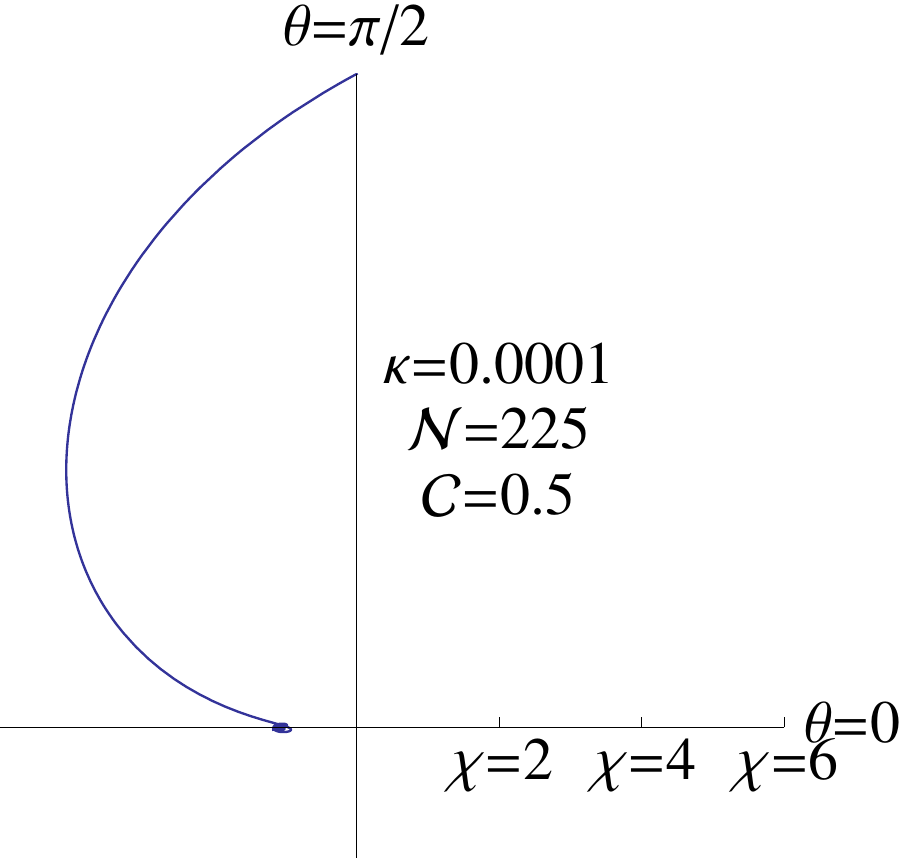}\\
  \end{tabular}
  \caption[Background trajectories]{Background inflationary trajectories with a flux parameter $g_{\rm{s}}M = 100\,\left[\mathcal{M}^{0} \right]$, inflaton mass $m_{0} = 5\,\left[\mathcal{M} \right]$ and the saturated Planck mass. Note $\mathcal{M}$ is the mass unit defined in Eq. (\ref{massunitdefinitionspin}). The horizontal axes denote $\chi \cos{\theta}$ while the vertical axes denote $\chi \sin{\theta}$. The value of $\kappa$ and $\mathcal{C} = c_{2} \kappa^{-4/3} [\mathcal{M}^{0}]$, are indicated. The same parameter sets are studied in \cite{spinflation}. }\label{backgroundreproduction}
\end{figure}

The system of Eqs. (\ref{equationofmotionchibackground}), (\ref{equationofmotionthetabackground}), (\ref{spinflationsaturatedfriedmann}) and (\ref{spinflationsaturatedhubblederivative}) can be solved numerically if we set the values of the parameters. We show the numerical results for the background trajectories with six different parameter sets following \cite{spinflation} in figure \ref{backgroundreproduction}. The flux parameter $g_{\rm{s}}M$ is set to 100 and the inflaton mass $m_{0}$ is set to 5 while we changed the values of $\kappa$ and $c_{2}$. Note that we assume $\chi_{\rm{UV}} = 10$ in this section. The trajectory starts with $\chi=10$ and $\theta=\pi/2$. The initial radial brane velocity is taken to vanish while the angular brane velocity is highly relativistic. Although the results are different from those in the published version of \cite{spinflation} due to a numerical problem, we confirmed some of their findings about the background dynamics \cite{Thanks:Ruth} as follows. 
\begin{itemize}
\item Decreasing the deformation parameter $\kappa$ slows down the brane and increases the number of e-folds. 
\item For all the parameter sets in figure \ref{backgroundreproduction}, regardless of the angular dependence or initial momenta, the brane rapidly becomes highly relativistic in the radial direction and makes its first sweep down the throat. 
\item Increasing the angular perturbation $c_{2}$ shifts the minimum of the potential. 
\end{itemize}

\begin{figure}
\centering
\includegraphics[width=12cm]{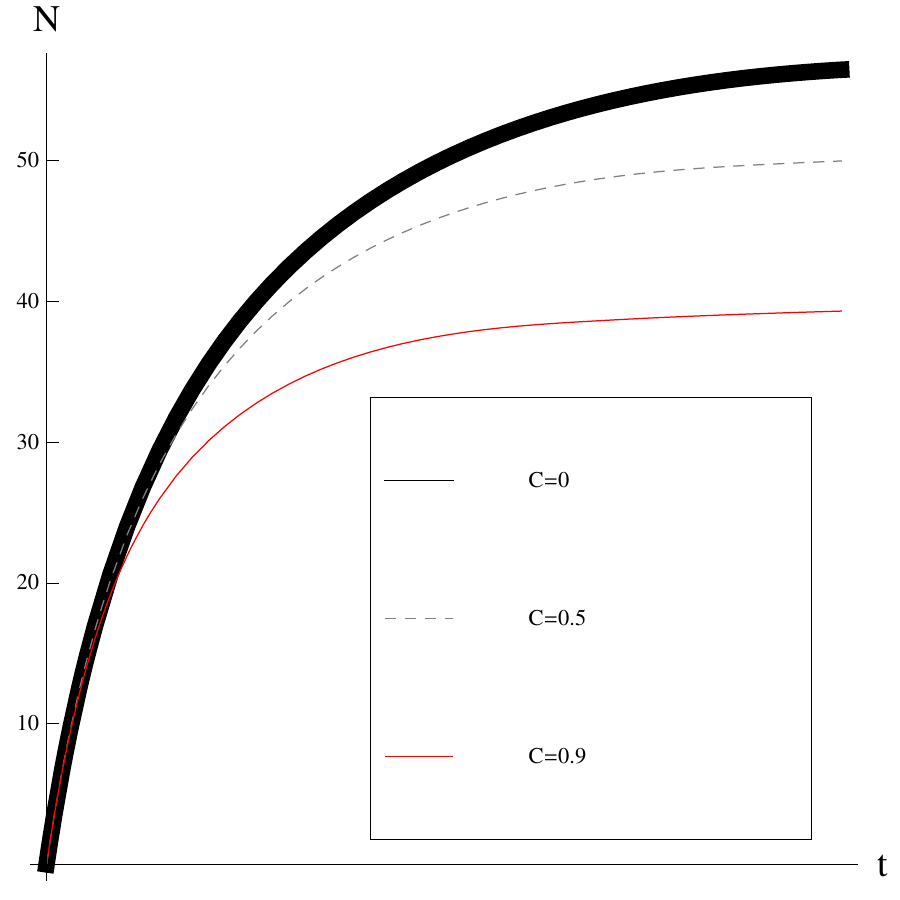}
\caption[Angular dependence]{The evolution of the number of e-folds along inflationary trajectories with $m_{0}=5\,\left[\mathcal{M} \right]$, $g_{\rm{s}}M=100\,\left[\mathcal{M}^{0} \right]$ and $\kappa=0.001\,\left[\mathcal{M}^{-3/2} \right]$.  Note $\mathcal{M}$ is the mass unit defined in Eq. (\ref{massunitdefinitionspin}). The thick black curve, the dashed grey curve and the red solid curve describe the trajectories with $c_{2}=0\,\left[\mathcal{M}^{-2} \right]$, $c_{2}=0.5 \kappa^{4/3}\,\left[\mathcal{M}^{-2} \right]$ and $c_{2}=0.9 \kappa^{4/3}\,\left[\mathcal{M}^{-2} \right]$ respectively. Note that $\mathcal{C} = c_{2} \kappa^{-4/3} [\mathcal{M}^{0}]$. The amount of inflation decreases as the angular dependence is increased.}{\label{angulardependence} }
\end{figure}

Below, we introduce the new findings in our numerical results. As we can see in the bottom pair of the plots in figure \ref{backgroundreproduction}, increasing the angular dependence changes the trajectory and decreases the number of e-folds. This is because the values of the slow-roll parameters increase due to the angular motion. Figure \ref{angulardependence} shows the decrease in the number of e-folds due to the increase in the angular dependence in more detail with different parameter sets. This is opposite to the finding about the angular dependence in \cite{spinflation}. The number of e-folds decreases by about 20 $\%$ when increasing $C=c_{2}\kappa^{-4/3}$ from 0.5 to 0.9 regardless of the value of $\kappa$ as shown in figure \ref{backgroundreproduction} and figure \ref{angulardependence}. We also checked that the effect on the number of e-folds stays the same even if we change the inflaton mass $m_{0}$. Therefore, the angular terms have some impacts on the background dynamics even though they are still sub-dominant.

In the top pair of the plots in figure \ref{backgroundreproduction}, we show the trajectories until the brane reaches the tip of the warped throat. With those parameter sets, the brane goes to the tip of the throat without reaching the minimum of the potential. This means that the radial coordinate $\chi$ overshoots the minimum in the radial direction before the angular coordinate $\theta$ reaches its minimum at $\theta=\pi$. This is because the velocity of the brane is high with a large deformation parameter as explained below. When the brane moves relativistically, the sound speed $c_{s}$ approaches unity and the speed limiting effect appears \cite{Underwood:2008}. Eq. (\ref{spinflationsoundspeedandfactor}) shows that the maximum speed of the brane is higher with a large deformation parameter $\kappa$ because of the factor $\kappa^{-8/3}$ in Eq. (\ref{warpfactor}). In other plots in figure \ref{backgroundreproduction}, the brane velocity is suppressed by the small deformation parameters and the brane moves slowly enough to settle at the minimum of the potential after some oscillations. The oscillations are smaller with a smaller deformation parameter because of the speed limiting effect. We checked that the increase in the number of e-folds during the oscillations around the minimum of the potential is negligible regardless of the choice of the parameter set and inflation occurs mainly in the initial sweep down the throat. 

\section{Linear perturbation}
\label{sec:linearperturbation}
In this section, we review the linear perturbation theory in DBI inflation models briefly. After introducing the conversion of the entropy perturbation into the curvature perturbation, we show the numerical results for such conversions with different initial conditions in the spinflation model introduced in subsection \ref{twofield}. Using the power spectrum of the curvature perturbation, we also show that the spinflation model with $\chi \gg 1$ is approximated by the simple model introduced in subsection \ref{singlefieldquadratic} when the coupling between the radial and angular fields is small around horizon crossing. 

\subsection{Linear perturbation theory}
\label{subsec:perturbationtheory}
For the calculations of the linear perturbations, ADM approach is used in \cite{Langlois:2008wt, Langlois:2008qf, Langlois:2008b} as
\begin{equation}\label{admfirstintroduction}
ds^{2}=-N^{2}dt^{2}+h_{ij}\left(dx^{i}+N^{i}dt \right)\left(dx^{j}+N^{j}dt\right),
\end{equation}
where $N$ is the lapse and $N^{i}$ is the shift vector. We perturb the components of the metric tensor as
\begin{equation}\label{mostgeneralperturbation}
N=1+\alpha,\,\, N_{i}=\partial_{i} \psi + \bar{N}_{i},\,\, h_{ij}= a^{2}\left(t\right)\left[(1-2A)\delta_{ij} + \bar{h}_{\vert ij} + \partial_{(i} v_{j)} + t_{ij}\right],
\end{equation}
where $\alpha$, $\psi$, $A$ and $\bar{h}$ are scalar perturbations, $\bar{N}_{i}$ and $v_{i}$ are vector perturbations, and $t_{ij}$ is a tensor perturbation with the Kronecker delta $\delta_{ij}$. Note that $_{\vert i}$ denotes the spatial covariant derivative with $a^{2}\delta_{ij}$. Because the scalar modes of the equations only contain the scalar modes of the original metric perturbations as long as the metric perturbations are contracted with the quantities which come from the background metric or the derivatives (see \cite{Kodama:1984} for details), we can consider the scalar perturbation separately from the vector and the tensor perturbations. We will work in the flat gauge where we set $A=0$ and $\bar{h}=0$. Then, we have the linear perturbations of the components of the metric tensor and the scalar fields as
\begin{equation}\label{allperturbations}
N=1+\alpha,\,\, N_{i}=\partial_{i} \psi,\,\, h_{ij}= a^{2}\left(t\right) \delta_{ij},\,\, \phi^{I}\left(t, \textbf{x}\right)=\bar{\phi}^{I}\left(t\right) + Q^{I}\left(t, \textbf{x}\right),
\end{equation}
where $Q^{I}$ are the scalar field perturbations. We define the adiabatic basis vector $\tilde{e}^{I}_{\sigma}$ as
\begin{equation}\label{newadiabaticbasis}
 \tilde{e}^{I}_{\sigma}=\frac{\sqrt{c_{s}}\dot{\phi}^{I}}{\sqrt{2X}},
\end{equation}
and define the entropy basis vector $\tilde{e}_{s}^{I}$ with the conditions
\begin{equation}\label{orthoconditions}
\begin{split}
 G_{IJ} \tilde{e}_{s}^{I} \tilde{e}_{s}^{J} &= \frac{1}{c_{s}},\\
 G_{IJ} \tilde{e}_{\sigma}^{I} \tilde{e}_{s}^{J} &=0.
\end{split}
\end{equation}
If we assume the relation
\begin{equation}
Q^{I} = \tilde{Q}_{\sigma} \tilde{e}^{I}_{\sigma} + \tilde{Q}_{s} \tilde{e}^{I}_{s},
\end{equation}
we obtain
\begin{equation}\label{newdefinitionofqsigma}
\tilde{Q}_{\sigma} \equiv \frac{G_{IJ} Q^{I} \tilde{e}^{J}_{\sigma}}{c_{s}},\,\,\,\tilde{Q}_{s} \equiv G_{IJ} Q^{I} \tilde{e}^{J}_{s} c_{s}. 
\end{equation}
We define the canonically normalised fields as
\begin{equation}
 v_{\sigma}=\frac{a}{c_{s}} \tilde{Q}_{\sigma}, \: \: \: \: \: v_{s}=\frac{a}{c_{s}} \tilde{Q}_{s}.
\label{relationofvandq}
\end{equation}
Then, the equations of motion for $v_{\sigma}$ and $v_{s}$ are obtained as \cite{Langlois:2008qf}
\begin{equation}
 v_{\sigma}'' - \xi v_{s}' + \left(c_{s}^{2} k^{2} - \frac{z''}{z} \right)v_{\sigma} - \frac{\left(z \xi \right)'}{z} v_{s} = 0,\label{equationofmotionone}
\end{equation}
\begin{equation}
 v_{s}'' + \xi v_{\sigma}' + \left(c_{s}^{2} k^{2} - \frac{\alpha''}{\alpha} + a^{2} \mu_{s}^{2} \right)v_{s} - \frac{z'}{z} \xi v_{\sigma} = 0,\label{equationofmotiontwo}
\end{equation}
where the prime denotes the derivative with respect to the conformal time $\tau$ and
\begin{equation}\label{xidefined}
 \xi \equiv \frac{a}{\dot{\sigma}} \left[\left(1+c_{s}^{2}\right) \tilde{P}_{,s} - c_{s}^{2} \dot{\sigma}^{2} \tilde{P}_{,\tilde{X} s} \right],
\end{equation}
\begin{equation}
 \mu_{s}^{2} \equiv -c_{s} \tilde{P}_{,ss} - \frac{1}{\dot{\sigma}^{2}} \tilde{P}_{,s}^{2} + 2 c_{2}^{2} \tilde{P}_{,\tilde{X} s} \tilde{P}_{,s},
\end{equation}
\begin{equation}
 z \equiv \frac{a \dot{\sigma}}{\sqrt{c_{s}} H}, \: \: \: \: \: \alpha \equiv a \frac{1}{\sqrt{c_{s}}},
\end{equation}
with
\begin{equation}\label{definitionsincludingsigmadotinchaptertwo}
 \dot{\sigma} \equiv \sqrt{2 X}, \:\:\: \tilde{P}_{s} \equiv \tilde{P}_{,I} e_{s}^{I} \sqrt{c_{s}}, \:\:\: \tilde{P}_{,\tilde{X} s} \equiv \tilde{P}_{,\tilde{X} I} e_{s}^{I} \sqrt{c_{s}}, \:\:\: \tilde{P}_{,ss} \equiv \left(\mathcal{D}_{I} \mathcal{D}_{J} \tilde{P} \right) e_{s}^{I} e_{s}^{J} c_{s},
\end{equation}
where $\mathcal{D}_{I}$ denotes the covariant derivative with respect to the field space metric $G_{IJ}$. With this field decomposition, the curvature perturbation is written as
\begin{equation}\label{chaptertwocurvaturewithnewbaseslinear}
 \mathcal{R} = \frac{H \sqrt{c_{s}}}{\dot{\sigma}} \tilde{Q}_{\sigma}. 
\end{equation}
From Eqs. (\ref{equationofmotionone}) and (\ref{equationofmotiontwo}), on small scales ($k \gg a H/c_{s}$), we can see that both the adiabatic mode $v_{\sigma}$ and the entropy mode $v_{s}$ propagate with the sound speed $c_{s}$ in the case of DBI inflation. If the trajectory is not curved significantly, the coupling $\xi/a H$ becomes much smaller than one. When the slow-roll parameters are much smaller than unity, the approximations $z''/z \simeq 2 / \tau^{2}$ and $\alpha'' / \alpha \simeq 2 / \tau^{2}$ hold. With those conditions, we can approximate Eqs. (\ref{equationofmotionone}) and (\ref{equationofmotiontwo}) as the Bessel differential equations. Then, the solutions with the Bunch-Davis vacuum initial conditions are given by
\begin{equation}
 v_{\sigma k} \simeq \frac{1}{\sqrt{2 k c_{s}}} e^{-i k c_{s} \tau} \left(1 - \frac{i}{k c_{s} \tau} \right),\label{dbisolone}
\end{equation}
\begin{equation}
 v_{s k} \simeq \frac{1}{\sqrt{2 k c_{s}}} e^{-i k c_{s} \tau} \left(1 - \frac{i}{k c_{s} \tau} \right),\label{dbisoltwo}
\end{equation}
when $\mu^{2}_{s} / H^{2}$ is negligible for the entropy mode \cite{Langlois:2008qf}. With the solution (\ref{dbisolone}), the curvature perturbation on super-horizon scales reads
\begin{equation}
 \mathcal{P}_{\mathcal{R}_{*}} = \frac{k^{3}}{2 \pi^{2}} \lvert \mathcal{R} \rvert^{2} = \frac{k^{3}}{2 \pi^{2}} \frac{\lvert v_{\sigma k} \rvert^{2}}{z^{2}} \simeq \left. \frac{H^{4}}{4 \pi^{2} \dot{\sigma}^{2}} \right\rvert_{*} \simeq \left. \frac{H^{2}}{8 \pi^{2} \epsilon c_{s}} \right\rvert_{*},
\label{curvaturepowerspectrumhorizon}
\end{equation}
where the subscript $*$ indicates that the corresponding quantity is evaluated at sound horizon crossing $k c_{s} = a H$. 

\subsection{Conversion of the entropy perturbation}\label{subsec:conversionentropy}
In this subsection, we first review the conversion mechanism of the entropy perturbation following \cite{Langlois:2008qf} with the different definitions of the adiabatic and entropy bases that are given in subsection \ref{subsec:perturbationtheory}. Then, we show the numerical results for such conversions in the spinflation model with different initial conditions. We define the entropy perturbation as
\begin{equation}\label{definedentropypert}
\mathcal{S} = \frac{H \sqrt{c_{s}}}{\dot{\sigma}} \tilde{Q}_{s},
\end{equation}
where $\tilde{Q}_{s}$ is defined in Eq. (\ref{newdefinitionofqsigma}). The equation of motion for the curvature perturbation is given by
\begin{equation}
\dot{\mathcal{R}} = \frac{\xi}{a}\mathcal{S} + \frac{H}{\dot{H}} \frac{c^{2}_{s} k^{2}}{a^{2}}\Psi,
\end{equation}
where $\Psi$ is the Bardeen potential defined as
\begin{equation}\label{afterdefinitionofbardeenpot}
\Psi = A + a H \left(\frac{\bar{h}'}{2} - a \psi \right),
\end{equation}
with $A$, $\bar{h}$ and $\psi$ in Eq. (\ref{mostgeneralperturbation}) and $\xi$ is defined in Eq. (\ref{xidefined}). We see that the entropy perturbation is the only source of the curvature perturbation on super-horizon scales $(c_{s} k / a H \ll 1)$. Therefore, on super-horizon scales, the equations of motion for the curvature perturbation and the entropy perturbation are given by
\begin{equation}\label{interactionevolutionequationentropy}
\dot{\mathcal{R}} \approx \alpha H \mathcal{S},\:\:\:\:\:\dot{\mathcal{S}} \approx \beta H \mathcal{S},
\end{equation}
where
\begin{equation}
 \alpha = \frac{\Xi}{c_{s} H},\label{alphamouse}
\end{equation}
\begin{equation}
 \beta = \frac{s}{2} - \frac{\eta}{2} - \frac{1}{3H^{2}}\left(\mu^{2}_{\rm{s}} + \frac{\Xi^{2}}{c^{2}_{\rm{s}}} \right),\label{betamouse}
\end{equation}
and
\begin{equation}\label{interactionredefinition}
 \Xi = \frac{c_{s}}{a}\xi.
\end{equation}
We can rewrite Eq. (\ref{interactionevolutionequationentropy}) as
\begin{equation}\label{transfermatrix}
\left( 
\begin{array}{@{\,}c@{\,}}
\mathcal{R}\\
\mathcal{S}
\end{array}
\right) = \left( 
\begin{array}{@{\,}cc@{\,}}
1&T_{\mathcal{R} \mathcal{S}}\\
0&T_{\mathcal{S} \mathcal{S}}\\
\end{array}
\right) \left( 
\begin{array}{@{\,}c@{\,}}
\mathcal{R}\\
\mathcal{S}
\end{array}
\right)_{*}
\end{equation}
where the subscript $*$ indicates that the corresponding quantity is evaluated at sound horizon crossing $kc_{s} = aH$ with
\begin{equation}
 T_{\mathcal{R} \mathcal{S}}(t_{*},t) = \int ^{t}_{t_{*}} \alpha(t')T_{\mathcal{S} \mathcal{S}}(t_{*},t')H(t')dt',\label{trs}
\end{equation}
\begin{equation}
 T_{\mathcal{S} \mathcal{S}}(t_{*},t) = \exp{\left(\int ^{t}_{t_{*}} \beta(t')H(t')dt'\right)}. \label{tss}
\end{equation}
Hence, the power spectrum of the curvature perturbation is given by
\begin{equation}\label{chaptertwotransferdefined}
\mathcal{P}_{\mathcal{R}} = \left(1 + T^{2}_{\mathcal{R} \mathcal{S}}\right) \mathcal{P}_{\mathcal{R}_{*}} = \frac{\mathcal{P}_{\mathcal{R}_{*}}}{\cos^{2}{\Theta}}.
\end{equation}
with
\begin{equation}
 \sin{\Theta} \equiv \frac{T_{\mathcal{R} \mathcal{S}}}{\sqrt{1 + T^{2}_{\mathcal{R} \mathcal{S}}}},\,\,\,\cos{\Theta} \equiv \frac{1}{\sqrt{1 + T^{2}_{\mathcal{R} \mathcal{S}}}},\label{sintheta}
\end{equation}
where $\mathcal{P}_{\mathcal{R}_{*}}$ is given by Eq. (\ref{curvaturepowerspectrumhorizon}) if the slow-roll parameters and $\xi/a H$ are much smaller than unity around the horizon crossing. 

\begin{figure}[htb]
  \centering
   \begin{tabular}{cc} 
    \includegraphics[width=70mm]{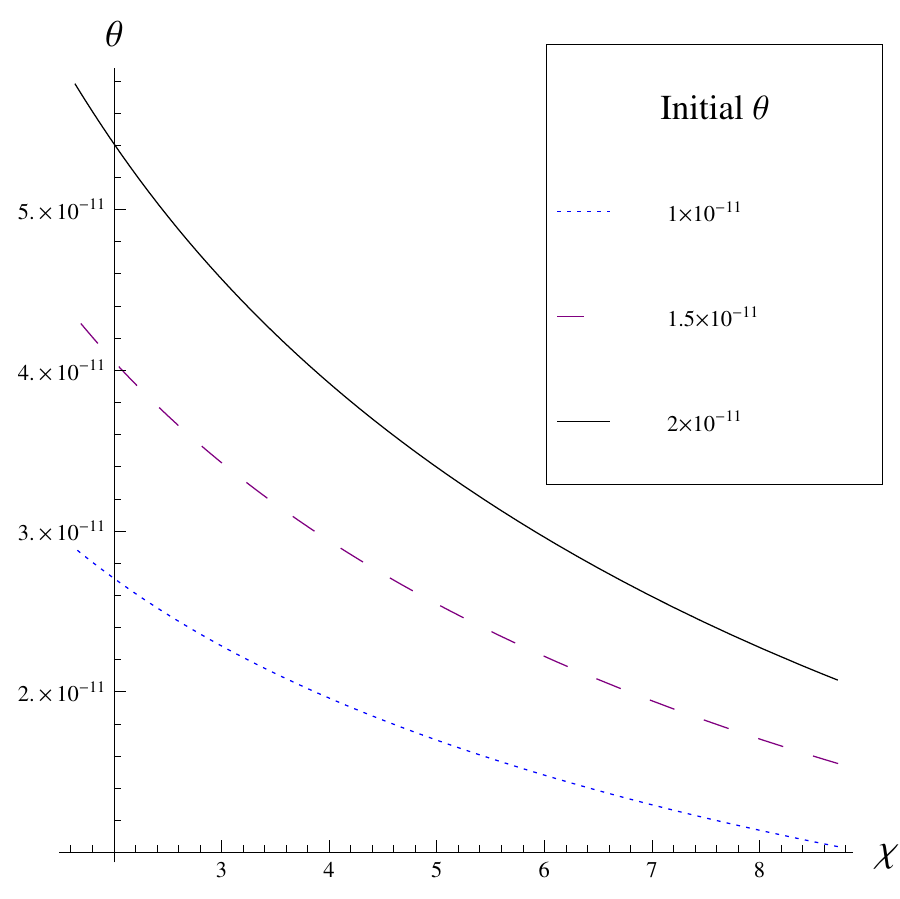}&
    \includegraphics[width=70mm]{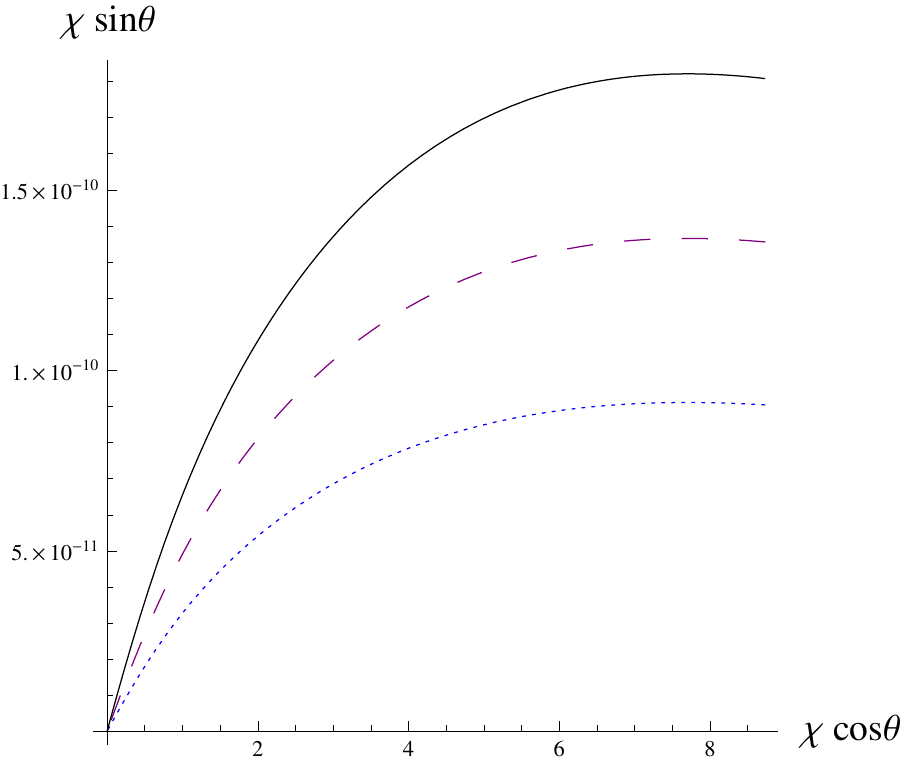}
  \end{tabular}
  \caption[Trajectories along the maximum]{Trajectories along the maximum of the potential at $\theta=0$. Left: trajectories in the $\chi$-$\theta$ plane. Right: trajectories in the phase space.}\label{trajectoriesalongmaximumspinflation}
\end{figure}

\begin{figure}[!htb]
  \centering
   \begin{tabular}{ccc} 
    \includegraphics[width=45mm]{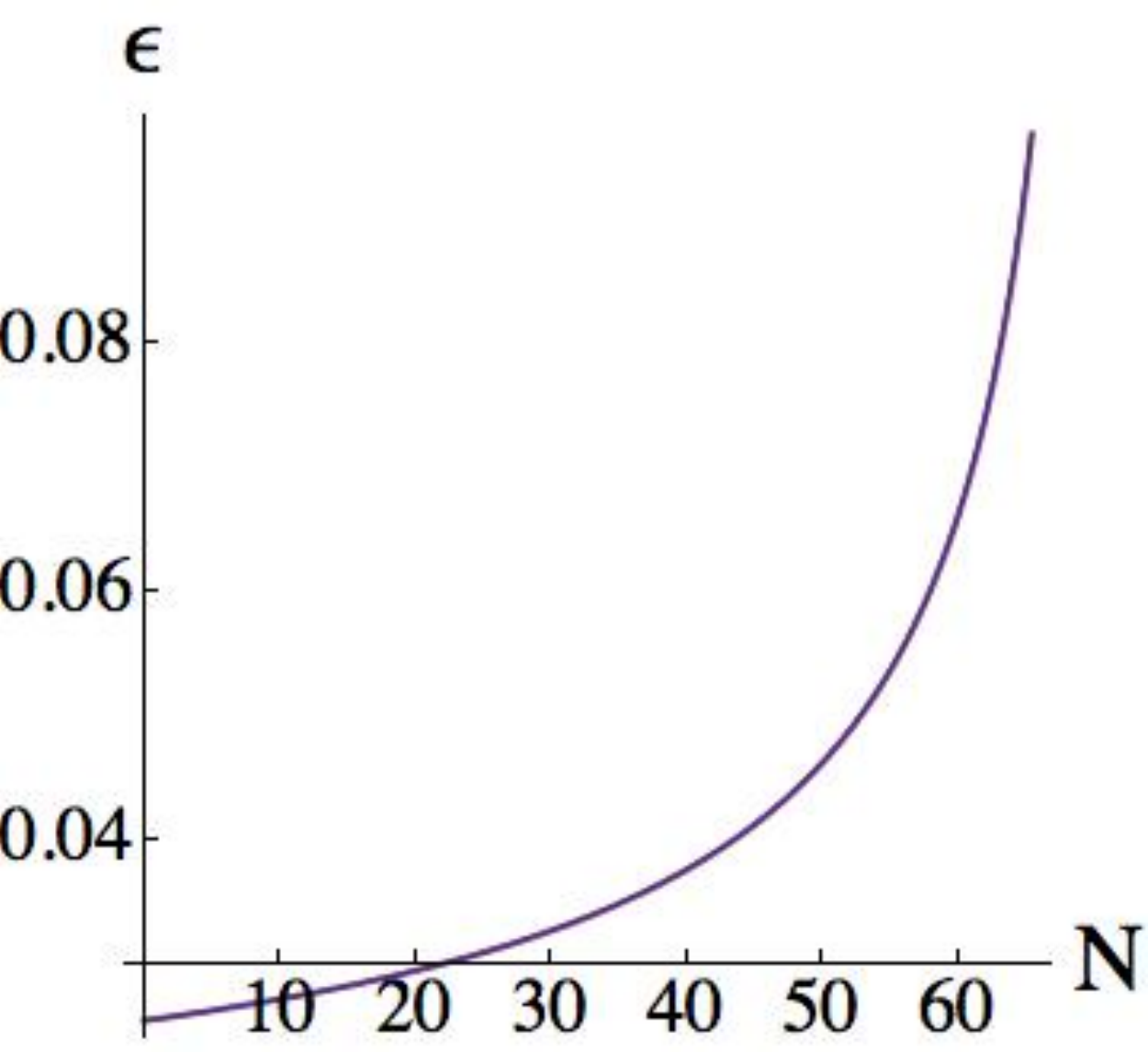}&
    \includegraphics[width=45mm]{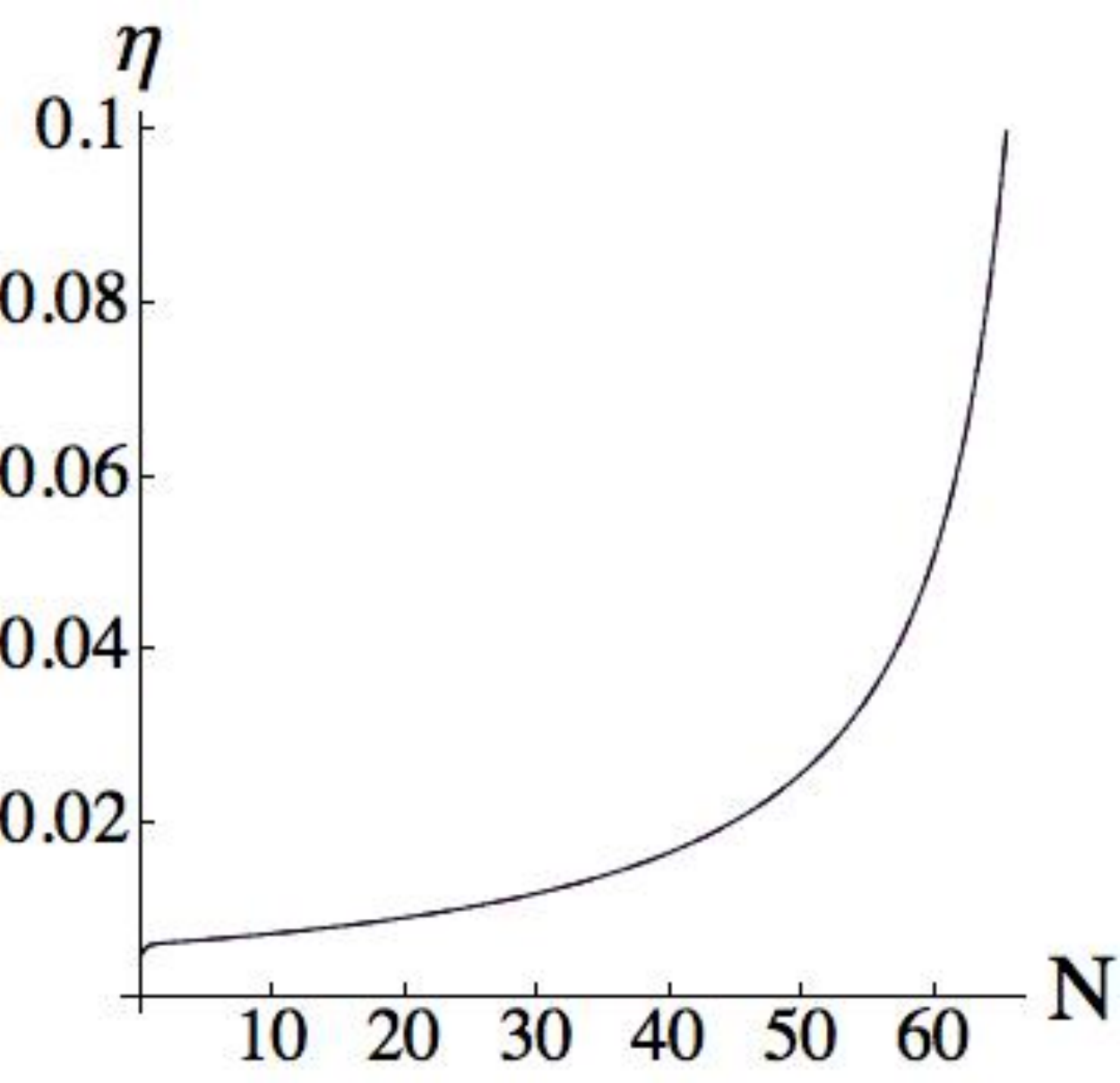}&
     \includegraphics[width=45mm]{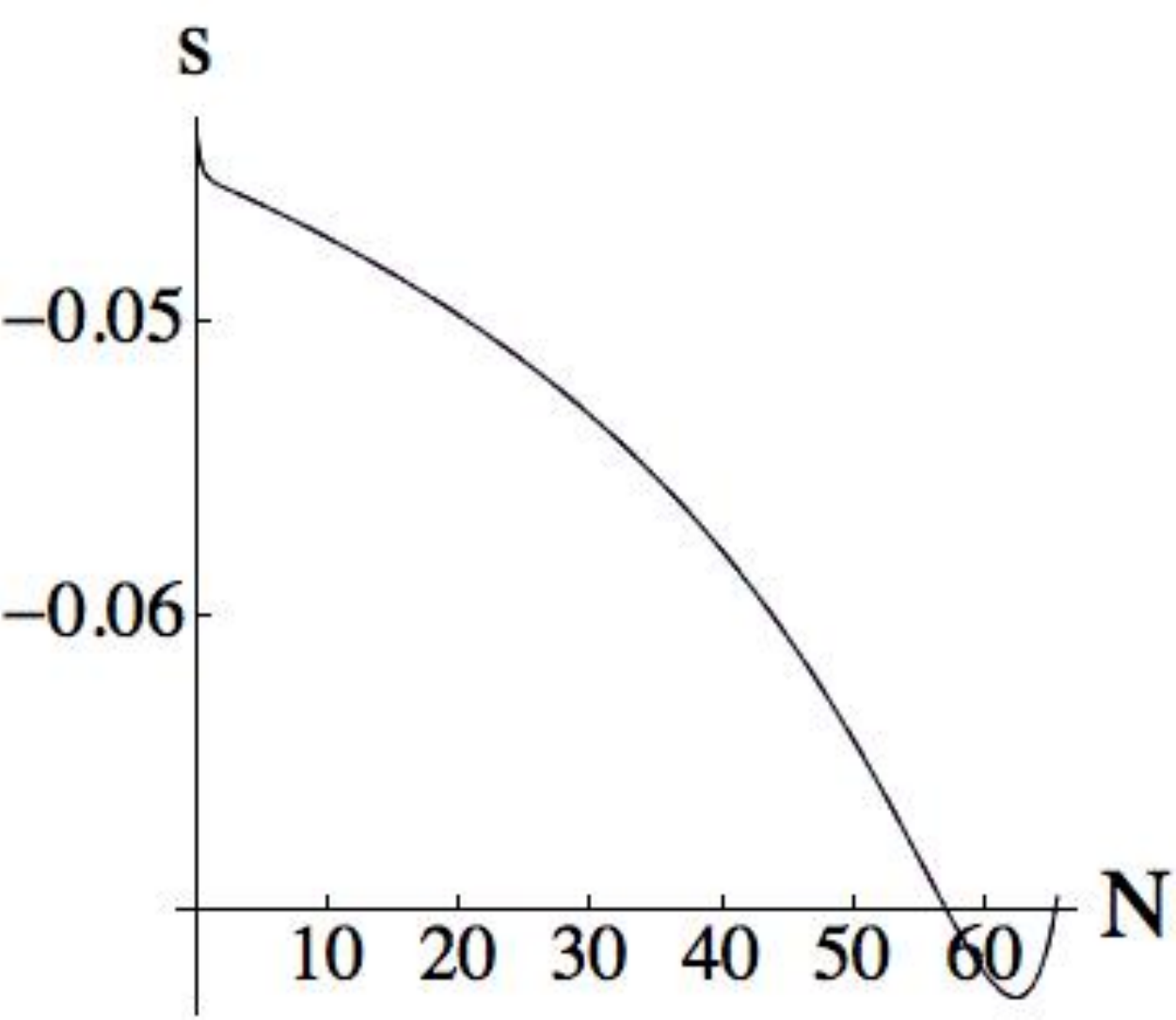}   
  \end{tabular}
  \caption[Slow-roll parameters]{Left: slow-roll parameter $\epsilon$. Middle: slow-roll parameter $\eta$. Right: slow-roll parameter $s$. All the slow-roll parameters behave in the same way for all the trajectories with $\theta = 1 \times 10^{-11}$, $1.5 \times 10^{-11}$ and $2 \times 10^{-11}$ because the displacements are small. The slow-roll approximation holds until the end of inflation.}\label{spinflationslowrollparametersmax}
\end{figure}

\begin{figure}[htb]
\centering
\includegraphics[width=10cm]{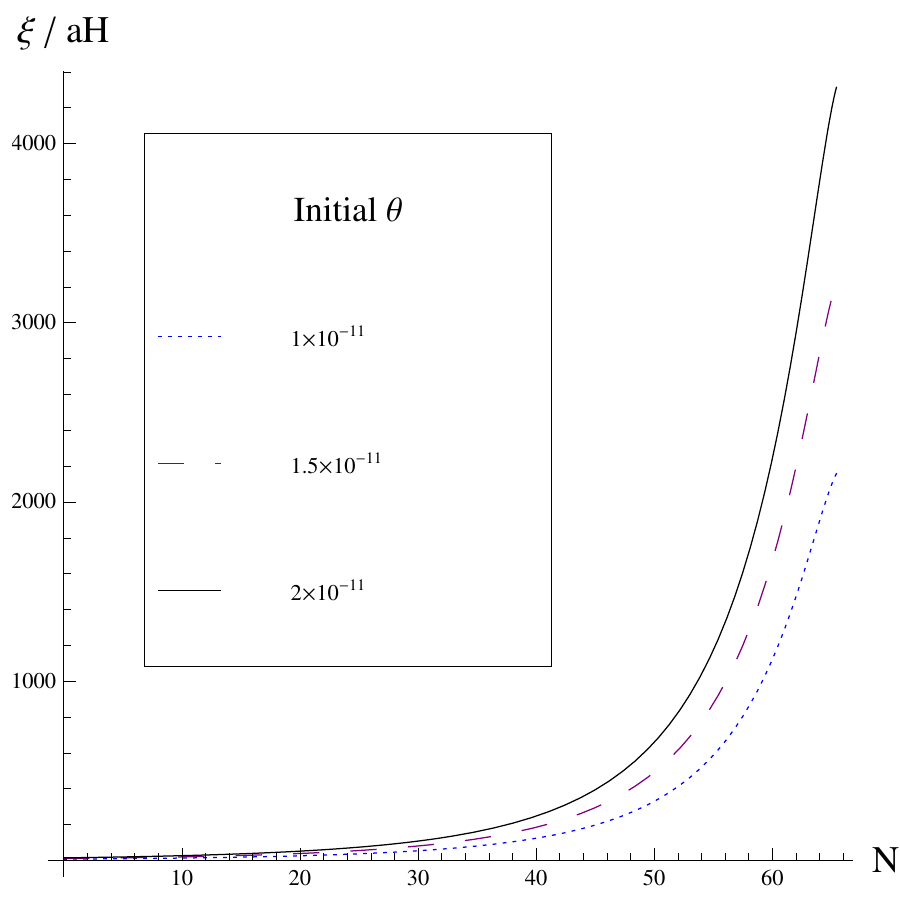}
\caption[Coupling along the maxmum]{The evolution of the coupling $\xi/a H$ in terms of the number of e-folds $N$ along the maximum of the potential in the angular direction.}{\label{couplingspinflationmax}}
\end{figure}

\begin{figure}
\centering
\includegraphics[width=10cm]{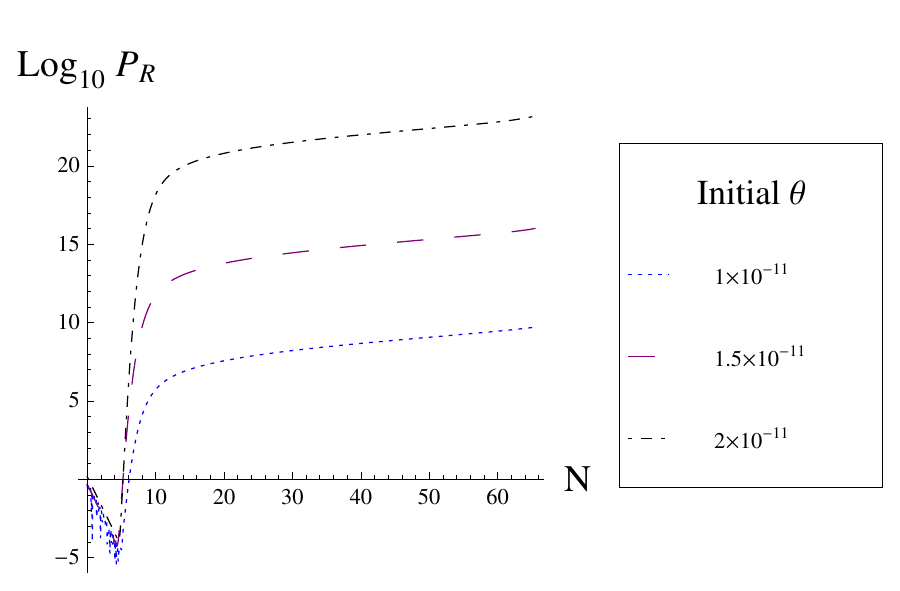}
\caption[Power spectrum of the curvature perturbation]{The evolution of the power spectrum of the curvature perturbation in terms of the number of e-folds $N$ along the maximum of the potential in the angular direction. The scale of interest exits the horizon around $N \sim 7$ for all the trajectories. }{\label{spinflationcurvaturemax}}
\end{figure}

We show how the power spectrum of the curvature perturbation is enhanced when we consider trajectories that start with slight deviations from the maximum of the potential in the angular direction in the spinflaion model. As shown in figure \ref{spinflationpotentialshape}, the potential has its minima in the angular direction at $\theta=\left(2N+1 \right)\pi$ where $N$ is an integer number. We set the initial conditions to $\left(\chi,\theta\right)=\left(9, \tilde{\delta}\theta\right)$ with $\tilde{\delta}\theta \ll 1$. The reason that we consider such trajectories is that the coupling $\xi/a H$ is small when the trajectory is a gentle curve around the minimum. If the coupling is not too large around horizon crossing, we can use analytic expressions that are useful to study the model, such as the solutions of the equations of motion for the linear perturbations (\ref{dbisolone}) and (\ref{dbisoltwo}), by starting the numerical calculations when the scale of interest is well within the horizon $k \gg a H/c_{s}$. Our numerical calculations show that the brane quickly becomes highly relativistic in the radial direction bending slowly towards the angular direction even if we set the initial velocity highly relativistic only in the angular direction. On the other hand, if the trajectory starts in the middle of the hill of the potential, it is bent towards the angular direction even if the initial velocity is only in the radial direction producing large coupling terms with $\xi/a H \gg 1$. Therefore, even though we have more conversion of the entropy perturbation to the curvature perturbation with a larger coupling, we study those cases in which the coupling is small and see how much conversion we have in those cases. We show the numerical results for three different trajectories with $\tilde{\delta}\theta = 1 \times 10^{-11}$, $\tilde{\delta}\theta = 1.5 \times 10^{-11}$ and $\tilde{\delta}\theta = 2 \times 10^{-11}$, which will be shown with a blue dotted line, a purple dashed line and a black solid line, respectively, in figures \ref{trajectoriesalongmaximumspinflation}, \ref{spinflationslowrollparametersmax}, \ref{couplingspinflationmax} and \ref{spinflationcurvaturemax}. 

In this case, the displacement from the maximum of the potential increases as inflation proceeds as shown in the left panel of figure \ref{trajectoriesalongmaximumspinflation}. The right panel shows that the brane goes to the tip of the throat $\chi = 0$ without reaching the minimum of the potential in the angular direction at $\theta = \pi$. The slow-roll parameters are shown in figure \ref{spinflationslowrollparametersmax}. Slow-roll approximation holds until the end of inflation. 

The coupling exhibits interesting behaviours in figure \ref{couplingspinflationmax}. The coupling increases as the number of e-folds increases. In addition to that, the difference between the trajectories also increases. This means that the coupling at the end of inflation could be large even if it is almost negligible around horizon crossing. Figure \ref{spinflationcurvaturemax} shows the numerical result for the power spectrum of the curvature perturbation. 

The values of $\xi/a H$ at horizon crossing are $10$, $15$, $22$ for the trajectories with $\tilde{\delta}\theta = 1 \times 10^{-11}$, $1.5 \times 10^{-11}$ and $2 \times 10^{-11}$ respectively. Therefore, we can no longer use the analytic solutions for the linear perturbations and the curvature power spectrum takes different values around horizon crossing with different trajectories. By starting the numerical calculations well within the horizon exits ($k \gg a H / c_{s}$), the values of the curvature power spectrum around horizon crossing are obtained numerically as $200$, $4\times10^{7}$ and $2 \times 10^{11}$ for the trajectories with $\tilde{\delta}\theta = 1 \times 10^{-11}$, $1.5 \times 10^{-11}$ and $2 \times 10^{-11}$, respectively, while the values of the final curvature power spectrum are $\sim10^{9}$,$\sim10^{16}$ and $\sim10^{24}$. 
Therefore, the values of $\cos^{2}{\Theta}$ are $\sim10^{-7}$, $\sim10^{-9}$ and $\sim10^{-15}$ for the trajectories with $\tilde{\delta}\theta = 1 \times 10^{-11}$, $1.5 \times 10^{-11}$ and $2 \times 10^{-11}$. In general, it is safe to assume that $\cos^{2}{\Theta}$ keeps decreasing as the amplitude of the displacement from the maximum increases. 

\subsection{Single field approximation of spinflation}
\label{subsec:approximationofspinflation}
The warp factor (\ref{warpfactor}) is approximated as \cite{spinflation, Klebanov:2000}
\begin{equation}\label{largechiwarpthroat}
h\left(\chi\right) \sim \frac{27}{8} \frac{\left(g_{\rm{s}}M\alpha' \right)^2}{r\left(\chi \right)^4} \left(\ln{\frac{r\left(\chi \right)^{3}}{\kappa^2}} + \ln{\frac{4\sqrt{2}}{3\sqrt{3}}} - \frac{1}{4} \right),
\end{equation}
for large $\chi > 1$. The angular term in the potential (\ref{potential}) is always smaller than the radial term because $c_{2}$ is smaller than $\kappa^{4/3}$ while $r\left(\chi \right)^{2}$ is of the order of $\kappa^{4/3}$. Also, the constant term $U_{0}$ is small by definition because the global minimum of the potential is at a point where $\chi \ll 1$ where both terms are negligible. Even though the angular term is not negligible in general, the radial term affects the dynamics dominantly when the motion of the brane is mainly in the radial direction. In such cases, the potential is approximated as
\begin{equation}\label{largechipotential}
V\left(\phi^{I} \right) \sim T_{3} U = T_{3} \left[ \frac{1}{2}m_{0}^{2} r\left(\chi\right)^{2} \right]. 
\end{equation}
When we compare the spinflation model in this section with the simple model in subsection \ref{singlefieldquadratic}, we can identify the fields $\chi$ and $\theta$ in this section with the dimensionless coordinates in Eq.~(\ref{scalarfieldwithbranetension}) as mentioned in section \ref{twofield}. Therefore, the canonical field $\phi$ with a mass dimension $\left[\mathcal{M} \right]$ is given by
\begin{equation}\label{spinexamplecanonicalfield}
\phi \left(\chi\right) = \sqrt{T_{3}} r \left(\chi\right) = \sqrt{T_{3}} \frac{\kappa^{2/3}}{\sqrt{6}}\int^{\chi}_{0}\frac{dx}{K\left(x\right)},
\end{equation}
where the dimensions of $T_{3}$ and $\kappa^{2/3}$ are $\left[\mathcal{M}^{4} \right]$ and $\left[\mathcal{M}^{-1} \right]$ respectively. 

Regarding the logarithmic dependence of $r$ as constant, the warp factor (\ref{largechiwarpthroat}) is approximated by Eq. (\ref{dbisingleexamplewarp}) with
\begin{equation}\label{largechilambda}
\begin{split}
\lambda &\equiv \frac{27\,T_{3}}{8} \left(g_{\rm{s}}M\alpha' \right)^{2} \left(\ln{\frac{r\left(\chi \right)^{3}}{\kappa^2}} + \ln{\frac{4\sqrt{2}}{3\sqrt{3}}} - \frac{1}{4} \right)\\ 
&= \frac{27}{64 \pi^3} g_{\rm{s}} M^{2} \left(\ln{\frac{r\left(\chi \right)^{3}}{\kappa^2}} + \ln{\frac{4\sqrt{2}}{3\sqrt{3}}} - \frac{1}{4} \right),
\end{split}
\end{equation}
where we used the relation \cite{spinflation}
\begin{equation}
T_{3}=\frac{1}{(2\pi)^{3}}\frac{1}{g_{\rm{s}}(\alpha')^{2}}.\label{definephi}
\end{equation}
Note that $f$ in Eq. (\ref{dbisingleexamplewarp}) is the rescaled warp factor $f = h/T_{3}$ with $h$ in Eq. (\ref{largechiwarpthroat}) as defined in Eq. (\ref{warpfactorandbranetension}). The potential (\ref{largechipotential}) is approximated by Eq. (\ref{dbisingleexamplepotential}) with
\begin{equation}
m \equiv m_{0}.
\end{equation}
We now show that the analytic formulae in subsection \ref{singlefieldquadratic} predict the numerical results with considerable accuracy. We consider a model with $g_{\rm{s}} = 1/2\pi \left[\mathcal{M}^{0} \right]$, $M = 1.2 \times 10^{6} \pi \left[\mathcal{M}^{0} \right]$, $m_{0} = 3 \times 10^{-3} \left[\mathcal{M} \right]$, $\kappa = 10^{-8} \left[\mathcal{M}^{-3/2} \right]$, $N = 1 \left[\mathcal{M}^{0} \right]$, $\alpha' = 10 \left[\mathcal{M}^{-2} \right]$, $\chi_{\rm{UV}} = 10 \left[\mathcal{M}^{0} \right]$ and $C = c_{2} \kappa^{-4/3} = 0.5 \left[\mathcal{M}^{0} \right]$. Note that all the quantities have the units associated with the mass unit $\mathcal{M}$ defined in Eq. (\ref{massunitdefinitionspin}). With those parameters, the mass unit $\mathcal{M}$ is defined by Eq. (\ref{massunitdefinitionspin}) as
\begin{equation}\label{massunitspinexample}
\mathcal{M} \simeq 0.0138 M_{\rm{P}}. 
\end{equation}
Therefore, for example, the string scale $1/\alpha \, \left[\mathcal{M}^{2} \right]$ in the Planck units is given by
\begin{equation}
\frac{\mathcal{M}^{2}}{\alpha' M^{2}_{\rm{P}}} = 1.91 \times 10^{-5} \left[M^{2}_{\rm{P}} \right]. 
\end{equation}

\begin{figure}[htp]
  \centering
   \begin{tabular}{cc} 
    \includegraphics[width=70mm]{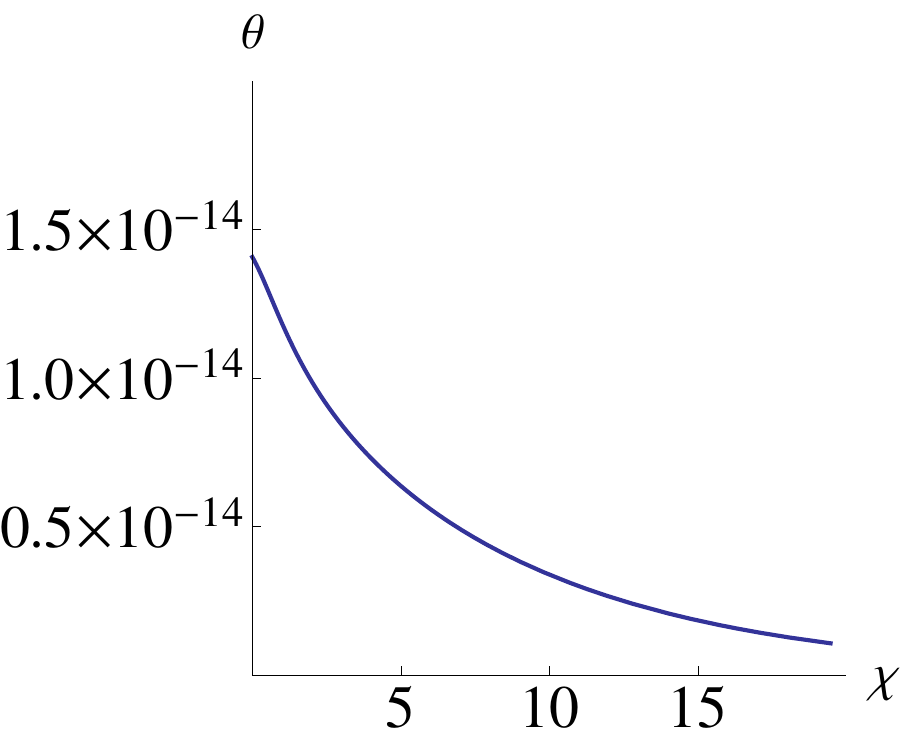}&
    \includegraphics[width=70mm]{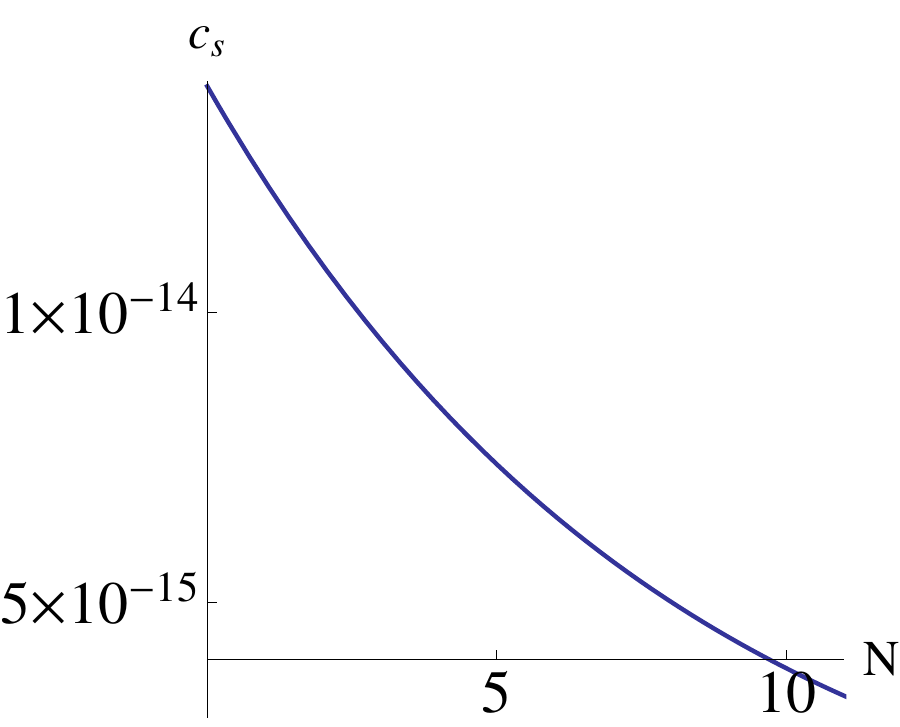}
  \end{tabular}
  \caption[Background trajectory and sound speed]{Left: background trajectory in the $\chi$-$\theta$ plane of the brane moving down the throat along the maximum of the potential in the angular direction with a small displacement from the maximum. Right: evolution of the sound speed in the early stage of inflation with respect to the number of e-folds.}\label{spinchithetaandsound}
\end{figure}

Figure \ref{spinchithetaandsound} shows the numerical results for the trajectory in the field space and the sound speed. The initial position of the brane is $\left(\chi,\,\theta \right) = \left(20, 10^{-15} \right)$. The initial velocity is only in the radial direction, even though we confirmed that the velocity becomes highly relativistic only in the radial direction, regardless of the initial velocity, when the trajectory is close to the maximum. In the left panel of figure \ref{spinchithetaandsound}, it is shown that the trajectory is bent towards the angular direction slowly and the deviation from the maximum of the potential in the angular direction becomes larger gradually. Below, we consider the perturbation that exits the horizon around $N \sim 2$. Using the numerical results, the value of the canonical field (\ref{spinexamplecanonicalfield}) around horizon crossing in the Planck units is
\begin{equation}
\frac{\phi \left(\chi \right)}{M_{\rm{P}}} = 4.50 \times 10^{-5} \frac{\mathcal{M}}{M_{\rm{P}}} = 6.22 \times 10^{-7} \left[M_{\rm{P}} \right], 
\end{equation}
where we used Eq. (\ref{massunitspinexample}). From Eq. (\ref{alltheresultsindbisingleexample}), the sound speed is given by
\begin{equation}\label{soundspeedpredictionspin}
c_{s} = \sqrt{\frac{3}{\lambda}} \frac{M_{\rm{P}}}{2 \bar{m}_{0} \mathcal{M}} \left(\frac{\phi}{M_{\rm{P}}}\right)^{2} \simeq 1.05 \times 10^{-14},
\end{equation}
with the dimensionless parameters $\bar{m}_{0} = m_{0}/\mathcal{M}$ and $\lambda = 5.78 \times 10^{11}$ that is given by Eq. (\ref{largechilambda}). In the right panel of figure \ref{spinchithetaandsound}, we see that the analytic formula (\ref{soundspeedpredictionspin}) predicts the sound speed around $N \sim 2$ with great accuracy.

\begin{figure}[htp]
  \centering
   \begin{tabular}{ccc} 
    \includegraphics[width=45mm]{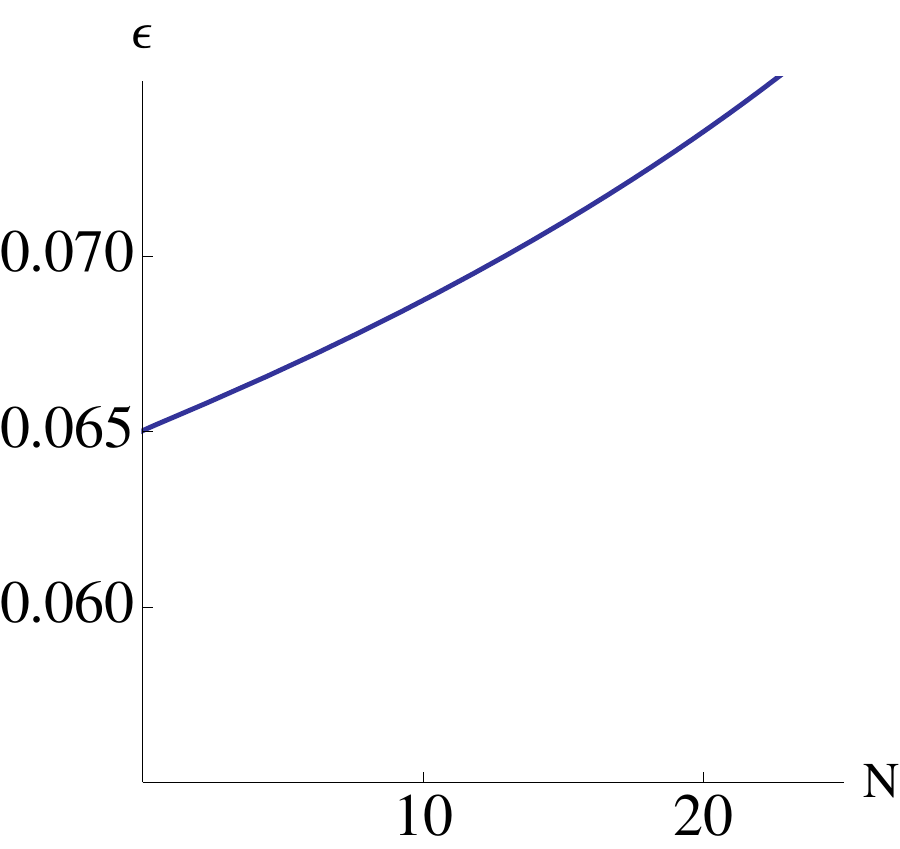}&
    \includegraphics[width=45mm]{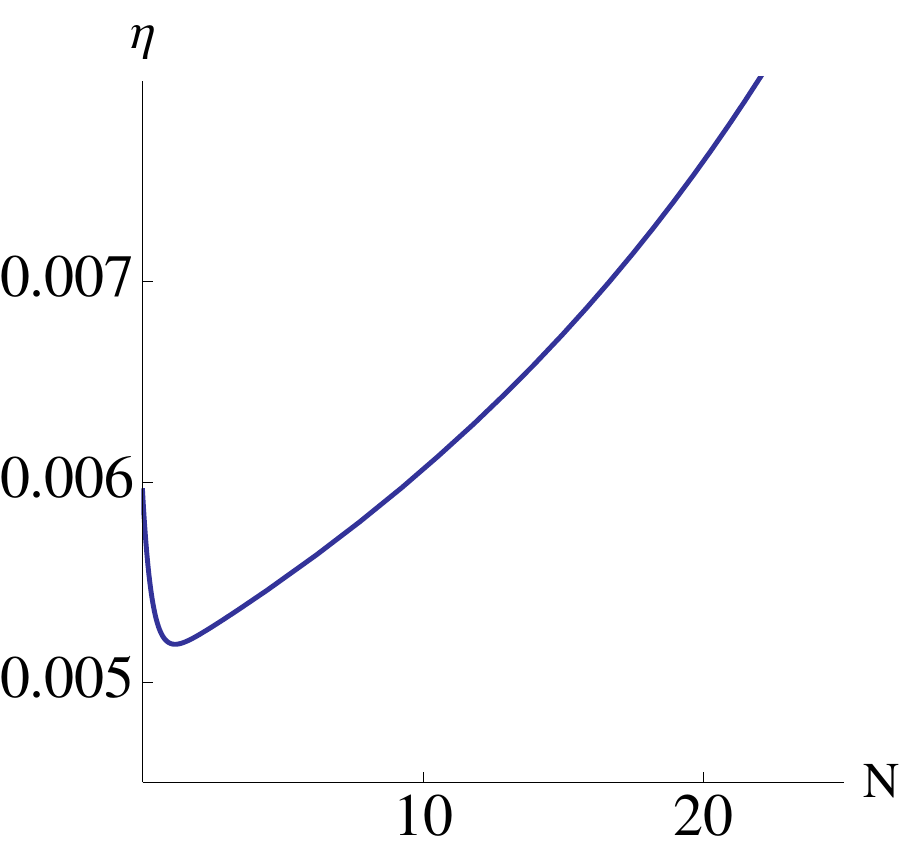}&
    \includegraphics[width=45mm]{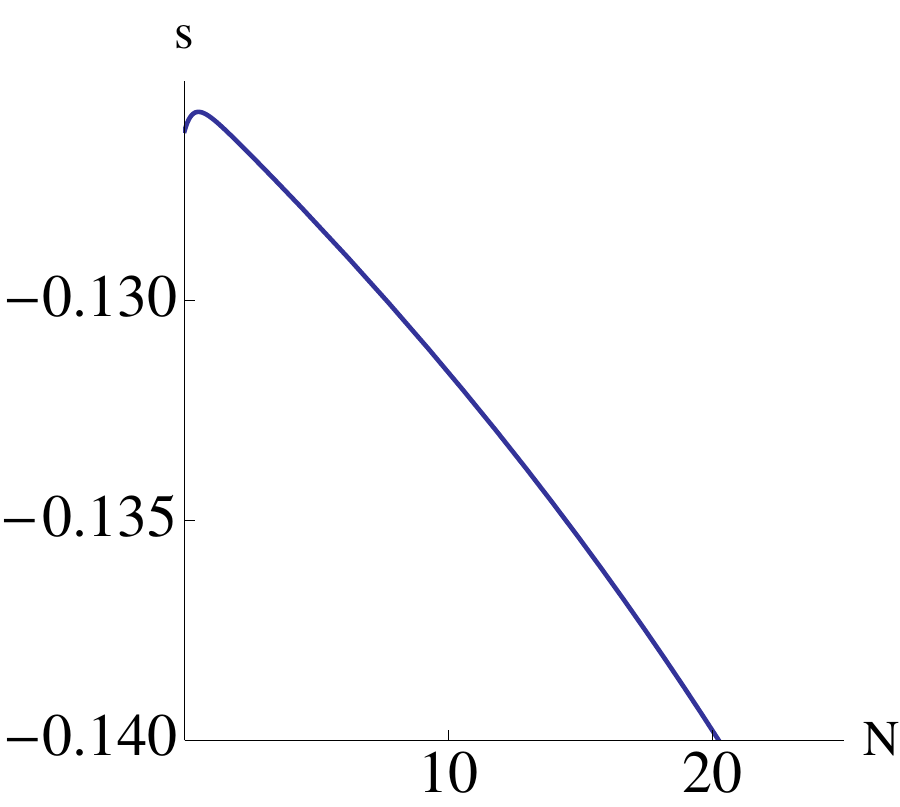}   
  \end{tabular}
  \caption[Slow-roll parameters]{Left: slow-roll parameter $\epsilon$. Middle: slow-roll parameter $\eta$. Right: slow-roll parameter $s$. All the horizontal axes denote the number of e-folds.}\label{spinslowrollexample}
\end{figure}

Figure \ref{spinslowrollexample} shows the behaviour of the slow-roll parameters. From Eq. (\ref{alltheresultsindbisingleexample}), the analytic prediction of the slow-roll parameter $\epsilon$ is given by
\begin{equation}\label{spinepsilonpredictionexample}
\epsilon = \sqrt{\frac{3}{\lambda}} \frac{M_{\rm{P}}}{\bar{m}_{0} \mathcal{M}} = 0.0545. 
\end{equation}
In the left panel of figure \ref{spinslowrollexample}, we see that the value of $\epsilon$ is predicted with the analytic formula (\ref{spinepsilonpredictionexample}) with around 20 $\%$ error. In the middle panel, $\eta$ is much smaller than $\epsilon$, whereas it is expected to vanish in Eq. (\ref{relationchapthreesimple}). The right panel shows that the second relation in Eq. (\ref{relationchapthreesimple}) holds as $s \simeq -2 \epsilon$. 

\begin{figure}[htp]
  \centering
   \begin{tabular}{ccc} 
    \includegraphics[width=45mm]{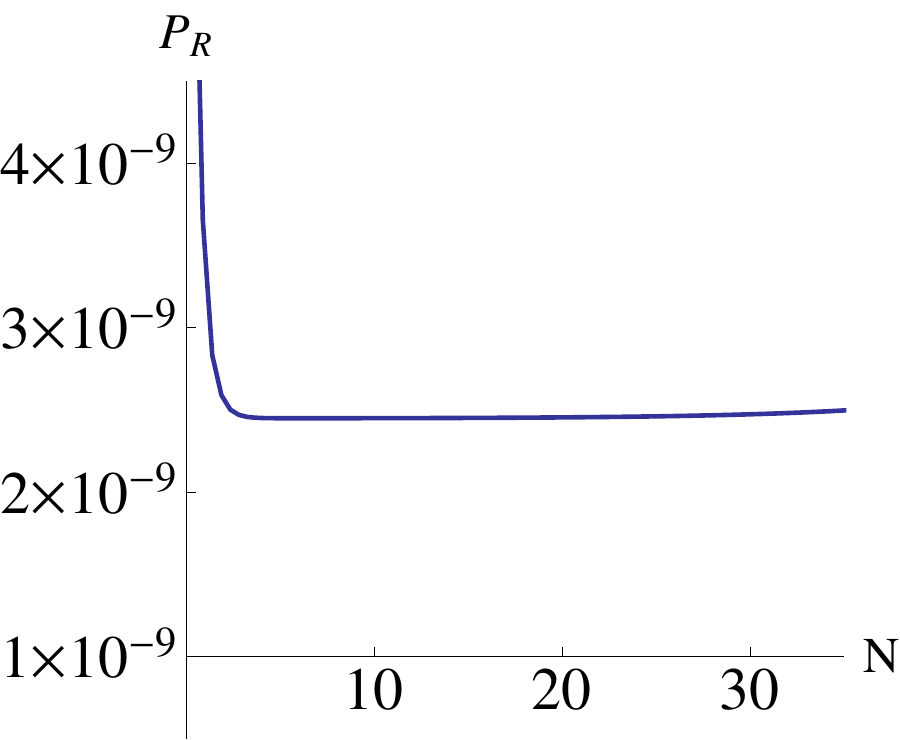}&
    \includegraphics[width=45mm]{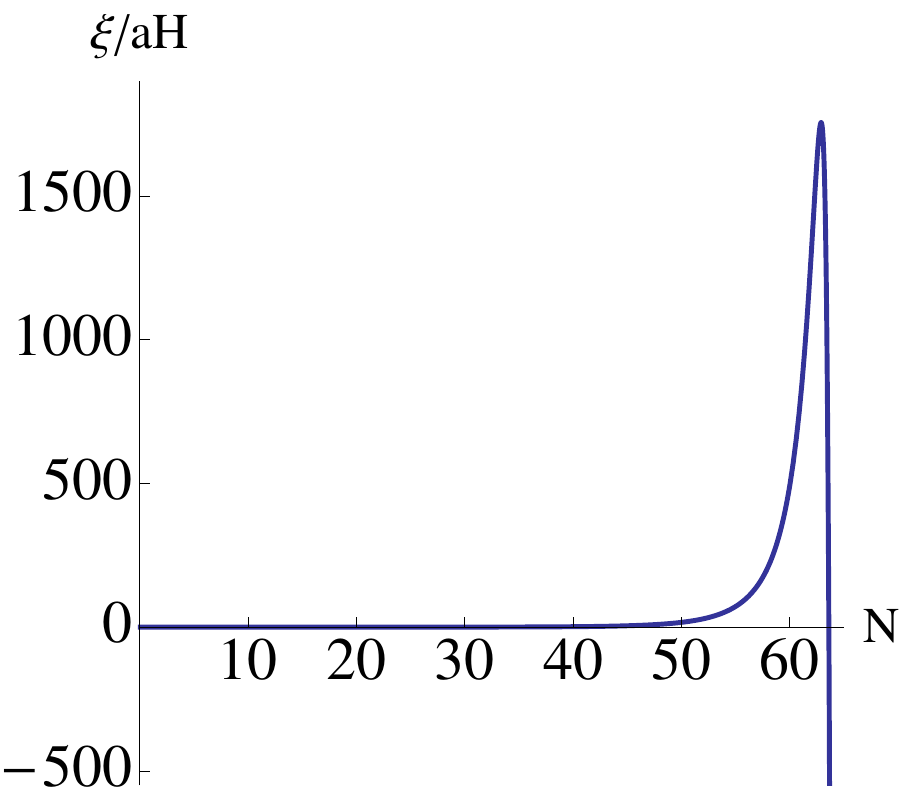}
    \includegraphics[width=45mm]{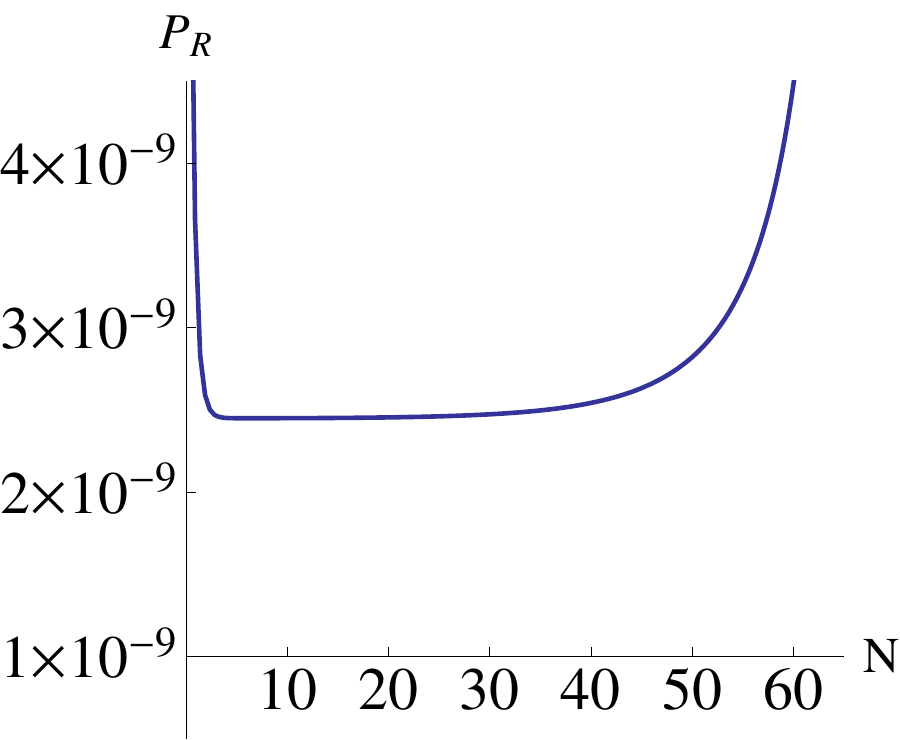}
  \end{tabular}
  \caption[Power spectrum of the curvature perturbation]{Left: curvature power spectrum in the early stage of inflation. Middle: coupling. It increases rapidly in the late stage of inflation. Right: behaviour of the curvature power spectrum until the end of inflation. }\label{spincurvatureandcouplingexample}
\end{figure}

As stated above, the trajectory is bent towards the angular direction gradually. As shown in the middle panel of figure \ref{spincurvatureandcouplingexample}, the coupling is negligible in the early stage of inflation and becomes larger rapidly in the late stage. Therefore, the curvature power spectrum $\mathcal{P}_{\mathcal{R}}$ is almost constant in the early stage of inflation as shown in the left panel of figure \ref{spincurvatureandcouplingexample}. Because this is the effective single field phase, it shows the same behaviour as the power spectrum of the curvature perturbation in the single field inflation models \cite{Nalson:2013, Nalson:2013b}. With Eq. (\ref{alltheresultsindbisingleexample}), the power spectrum of the curvature perturbation (\ref{curvaturepowerspectrumhorizon}) reads
\begin{equation}\label{curvaturespectrumexampledbisingle}
\mathcal{P}_{\mathcal{R}} = \frac{1}{8 \pi^{2} M^{2}_{P}}\frac{H^{2}}{c_{\rm{s}}\epsilon} = \frac{1}{4 \pi^{2} \epsilon^{4} \lambda}.
\end{equation}
Therefore, the value of the power spectrum is analytically predicted as
\begin{equation}\label{curvatureexamplespin}
\mathcal{P}_{\mathcal{R}_{*}} = \frac{1}{4 \pi^{2} \epsilon^4 \lambda} \simeq 2.42 \times 10^{-9},
\end{equation}
where we used $\epsilon = 0.065$ and $\lambda = 5.78 \times 10^{11}$ that is obtained with Eq. (\ref{largechilambda}). In the left panel of figure \ref{spincurvatureandcouplingexample}, it is shown that the approximated analytic formula (\ref{curvatureexamplespin}) predicts the power spectrum of the curvature perturbation with great accuracy when the brane has the effective single field dynamics. In the right panel, we see that the curvature power spectrum is enhanced in the late stage of inflation because of the coupling between the adiabatic and entropy perturbations. Even though it is enhanced only by the factor of 2 with those parameters and the initial conditions, the conversion of the entropy perturbation into the curvature perturbation becomes larger if we make the initial displacement from the angular maximum larger. 

\section{Non-Gaussianity}\label{sec:nongaussianity}
In this section, we introduce the equilateral non-Gaussianity. We also review the microphysical constraint that excludes single field DBI inflation models and show that multi-field DBI inflation models have the possibility of satisfying the constraint with the conversion mechanism introduced in subsection \ref{subsec:conversionentropy}. Finally, using the observables including the equilateral non-Gaussianity, we show that the spinflation model introduced in subsection \ref{twofield} is excluded by the Planck satellite observations even with the conversion mechanism in the regime where the approximation shown in subsection \ref{subsec:approximationofspinflation} holds. 
\subsection{Equilateral non-Gaussianity}
In this subsection, we review the non-Gaussianity parameter $f_{\rm{NL}}^{equil}$ in single field DBI inflation models \cite{Arroja:2008yy, Creminelli:2006, Maldacena:2005, Chen:2007}. Then, we introduce $f_{\rm{NL}}^{equil}$ in multi-field DBI inflation models following \cite{Langlois:2008qf}. The parameter $f_{\rm{NL}}^{equil}$ quantifies the bispectrum of the curvature perturbation as
\begin{equation}\label{fnlequilateraldefinition}
\begin{split}
&\left<\Omega \left\lvert \mathcal{R}\left(0, \textbf{k}_{\textbf{1}} \right) \mathcal{R}\left(0, \textbf{k}_{\textbf{2}} \right) \mathcal{R}\left(0, \textbf{k}_{\textbf{3}} \right) \right\lvert \Omega\right>^{\left(3\right)}\\
&= - (2\pi)^{7} \delta^{\left(3\right)}\left(\textbf{k}_{\textbf{1}}+\textbf{k}_{\textbf{2}}+\textbf{k}_{\textbf{3}}\right) \left(\frac{3}{10} f_{\rm{NL}}^{equil} \left(\mathcal{P}_{\mathcal{R}_{*}} \right)^{2} \right) \frac{\sum_{i}k^{3}_{i}}{\Pi_{i}k^{3}_{i}},
\end{split}
\end{equation}
where $\mathcal{R}\left(\tau, \textbf{k}_{\textbf{i}} \right)$ is the Fourier component of the curvature perturbation with the wave vector $\textbf{k}_{\textbf{i}}$ $\left(i = 1,2,3\right)$. Note that we take the conformal time $\tau \approx - (aH)^{-1}$ to be 0 at the end of inflation. We can derive the bispectrum of the curvature perturbation in the single field DBI inflation using the in-in formalism as \cite{Chen:2007, Weinberg:2005}
\begin{equation}\label{singledbithreepoint}
\begin{split}
&\left<\Omega \left\lvert \mathcal{R}\left(0, \textbf{k}_{\textbf{1}} \right) \mathcal{R}\left(0, \textbf{k}_{\textbf{2}} \right) \mathcal{R}\left(0, \textbf{k}_{\textbf{3}} \right) \right\lvert \Omega\right>^{\left(3\right)}\\
& = - \left(2\pi\right)^{7} \delta^{\left(3\right)}\left(\textbf{k}_{\textbf{1}}+\textbf{k}_{\textbf{2}}+\textbf{k}_{\textbf{3}}\right) \left(\mathcal{P}_{\mathcal{R}}\right)^{2} \frac{1}{\Pi_{i}k^{3}_{i}} \mathcal{A}^{DBI} \left(k_{1},k_{2},k_{3}\right), 
\end{split}
\end{equation}
where
\begin{equation}\label{singledbia}
\begin{split}
\mathcal{A}^{DBI} &= \frac{1-c_{s}^{2}}{c_{s}^{2}} \left(-\frac{1}{K}\sum_{i>j} k^{2}_{i}k^{2}_{j} + \frac{1}{2K^{2}}\sum_{i \neq j} k^{2}_{i}k^{3}_{j} + \frac{1}{8} \sum_{i} k^{3}_{i} \right)\\
& + \frac{\epsilon}{c^{2}_{s}}\left(- \frac{1}{8} \sum_{i} k^{3}_{i} + \frac{1}{8}\sum_{i \neq j} k^{2}_{i}k^{2}_{j} + \frac{1}{K}\sum_{i>j} k^{2}_{i}k^{2}_{j} \right) + \frac{\eta}{c^{2}_{s}}\left(\frac{1}{8} \sum_{i} k^{3}_{i} \right)\\
& + \frac{s}{c^{2}_{s}} \left(- \frac{1}{4} \sum_{i} k^{3}_{i} - \frac{1}{K}\sum_{i>j} k^{2}_{i}k^{2}_{j} + \frac{1}{2K^{2}}\sum_{i \neq j} k^{2}_{i}k^{3}_{j} \right),
\end{split}
\end{equation}
The bispectrum of the curvature perturbation (\ref{singledbithreepoint}) has its maximum at the equilateral configuration $k_{1} \sim k_{2} \sim k_{3}$ as shown in \cite{Creminelli:2006}. Even though the ``bispectrum" of the curvature perturbation (\ref{fnlequilateraldefinition}) is not the same with the actual bispectrum (\ref{singledbithreepoint}), it is defined so that it has the same value as the actual bispectrum at the equilateral configuration where it has its maximum. By setting $k_{1} = k_{2} = k_{3} = \tilde{k}$,  we have
\begin{equation}\label{limitequilateral}
\mathcal{A}^{DBI} = -\frac{7}{24} \left(\frac{1}{c^{2}_{s}} - 1 \right) \tilde{k}^{3},
\end{equation}
at the leading order. From Eqs. (\ref{singledbithreepoint}) and (\ref{limitequilateral}), in this limit, we have
\begin{equation}\label{equilateralfnlderivation}
\begin{split}
&\left<\Omega \left\lvert \mathcal{R}\left(0, \textbf{k}_{\textbf{1}} \right) \mathcal{R}\left(0, \textbf{k}_{\textbf{2}} \right) \mathcal{R}\left(0, \textbf{k}_{\textbf{3}} \right) \right\lvert \Omega\right>^{\left(3\right)}\\
&= - (2\pi)^{7} \delta^{\left(3\right)}\left(\textbf{k}_{\textbf{1}}+\textbf{k}_{\textbf{2}}+\textbf{k}_{\textbf{3}}\right) \left(-\frac{7}{24} \left(\frac{1}{c^{2}_{s}} - 1 \right) \frac{\tilde{k}^{3}}{\Pi_{i}k^{3}_{i}} \right) \left(\frac{H^{2}}{8 \pi^2 \epsilon c_{s}} \right)^{2}\\
&= - (2\pi)^{7} \delta^{\left(3\right)}\left(\textbf{k}_{\textbf{1}}+\textbf{k}_{\textbf{2}}+\textbf{k}_{\textbf{3}}\right) \left(-\frac{7}{72} \left(\frac{1}{c^{2}_{s}} - 1 \right) \frac{\sum_{i}k^{3}_{i}}{\Pi_{i}k^{3}_{i}} \right) \left(\mathcal{P}_{\mathcal{R}_{*}} \right)^{2}\\
&= - (2\pi)^{7} \delta^{\left(3\right)}\left(\textbf{k}_{\textbf{1}}+\textbf{k}_{\textbf{2}}+\textbf{k}_{\textbf{3}}\right) \frac{3}{10} \left(-\frac{35}{108} \left(\frac{1}{c^{2}_{s}} - 1 \right) \frac{\sum_{i}k^{3}_{i}}{\Pi_{i}k^{3}_{i}} \right) \left(\mathcal{P}_{\mathcal{R}_{*}} \right)^{2}.
\end{split}
\end{equation}
Comparing Eq. (\ref{fnlequilateraldefinition}) with Eq. (\ref{equilateralfnlderivation}), we obtain
\begin{equation}\label{fnlequilateralfinal}
f_{\rm{NL}}^{equil} = -\frac{35}{108} \left(\frac{1}{c^{2}_{s}} - 1 \right). 
\end{equation}
The non-Gaussianity parameter $f_{\rm{NL}}^{equil}$ in two-field DBI inflation models is derived in the small sound speed limit $c_{s} \ll 1$ in the slow-roll approximation as
\begin{equation}\label{dbimultiequilateralfinalresult}
f_{\rm{NL}}^{equil} = - \frac{35}{108} \left(\frac{1}{c^{2}_{s}} - 1\right) \frac{1}{1 + T_{\mathcal{R}\mathcal{S}}^{2}} = - \frac{35}{108} \left(\frac{1}{c^{2}_{s}} - 1\right) \cos^{2}{\Theta},
\end{equation}
where $T_{\mathcal{R}\mathcal{S}}$ and $\cos{\Theta}$ are defined in Eqs. (\ref{transfermatrix}) and (\ref{sintheta}) respectively.

\subsection{Microphysical constraint}\label{sebsec:microphysical}
In this subsection, we introduce the microphysical constraint that strongly disfavours single field DBI inflation models. We also show how multi-field DBI inflation models have the possibility of satisfying this constraint. Baumann and McAllister (2006) derived an upper bound on the tensor-to-scalar ratio by analysing the higher dimensional geometry that can be approximated with geometry $AdS_{5} \times X_{5}$ \cite{Baumann:2007}. In DBI inflation models, the scalar fields describe the position of the brane in the higher dimensional manifold. Because the volume of such a higher dimensional manifold is finite, the variation of the inflaton field during the observable inflation $\Delta \phi_{*} = \sqrt{T_{3}} \Delta \chi_{*}$ must be finite as well. This is simply because the brane cannot move across an infinite distance within the higher dimensional manifold whose volume is finite. Therefore, the upper limit of the variation of the inflaton field during the observable inflation is derived as \cite{Baumann:2007, Lidsey:2007}
\begin{equation}\label{upperboundofvariationofinflaton}
 \left(\frac{\Delta \phi_{*}}{M_{P}}\right)^{6} < \frac{\pi^{3}}{16 \rm{Vol}(X_{5})}r^{2}_{*}\mathcal{P}_{\mathcal{R}}\left(1+ \frac{1}{3f_{\rm{NL}}^{equil}} \right).
\end{equation}
where $\rm{Vol}(X_{5})$ is the dimensionless volume of the space $X_{5}$.This condition weakly depends on the non-Gaussianity parameter in the case that $f_{\rm{NL}}^{equil} > 5$ which is still compatible with the Planck satellite observations. Therefore, we neglect the factor with $f_{\rm{NL}}^{equil}$ in the condition (\ref{upperboundofvariationofinflaton}). We usually expect $\rm{Vol}(X_{5}) = \mathcal{O}(\pi^{3})$. Using the Lyth bound \cite{Lyth:2005}
\begin{equation}
 \frac{1}{M^{2}_{P}}\left(\frac{\Delta \phi}{\Delta N}\right)^{2} = \frac{r}{8},\label{lythbound}
\end{equation}
the upper bound on the variation of the inflaton field (\ref{upperboundofvariationofinflaton}) is rewritten as the upper bound on the tensor-to-scalar ratio as
\begin{equation}
r_{*} < 10^{-7}, \label{rcondition}
\end{equation}
assuming that the minimum number of e-folds that could be probed by observation is $\Delta N \sim 1$ and the Planck normalisation $\mathcal{P}_{\mathcal{R}} = 2.23 \times 10^{-9}$ \cite{Ade:2013c}. Secondly, the lower bound on the tensor-to-scalar ratio in the single field UV DBI inflation is derived in the following way. The general relation between the spectral index for the curvature perturbation and $f^{\rm{equil}}_{\rm{NL}}$ in the multi-field DBI inflation is given by \cite{Langlois:2009}
\begin{equation}
 1 - n_{\rm{s}} \simeq \frac{\sqrt{3 \lvert f^{\rm{equil}}_{\rm{NL}} \rvert}r}{4 \cos^{3}{\Theta}} - \frac{\dot{f}}{Hf} + \alpha_{*} \sin{2 \Theta} + 2 \beta_{*} \sin^{2}{\Theta},
\label{multispectral}
\end{equation}
where $f$ is the warp factor defined in Eq. (\ref{warpfactorandbranetension}) and $\sin{\Theta}$ is defined in Eq. (\ref{sintheta}). Note that a term proportional to $c_{s}^{2} s_{*}$ is neglected because we assume that both $c_{s}$ and $s_{*}$ are small. For the single field UV DBI inflation, we have $\dot{f} > 0$ and $\Theta = 0$. Therefore, we have
\begin{equation}
 r > \frac{4}{\sqrt{3 \lvert f^{\rm{equil}}_{\rm{NL}} \rvert}} \left( 1 - n_{s} \right) \: \: \: \: \: \: \: \: \: \: (\rm{single \:\: field}),
\end{equation}
from Eq. (\ref{multispectral}). The amplitude of the equilateral non-Gaussianity is constrained as \cite{Ade:2013c}
\begin{equation}\label{fnlplanckobservationvaluesixtyeight}
f_{\rm{NL}}^{equil} = -42 \pm 75,
\end{equation}
and the best-fit value for the specrtral index is $n_{s} \simeq 0.96$ from the Planck satellite observation. From those values, we can obtain the lower bound on the tensor-to-scalar ratio as
\begin{equation}
 r \gtrsim 8.4 \times 10^{-3}.
\label{lower}
\end{equation}
The lower bound (\ref{lower}) is not compatible with the upper bound (\ref{rcondition}). This is why single field UV DBI inflation is disfavoured by observation.

These constraints are relaxed when we consider multi-field models. The upper bound is relaxed because we have angular directions and the field variation is not only determined by the radial coordinate. More importantly, the lower bound is relaxed significantly because the last two terms in Eq. (\ref{multispectral}) become important if there is a transfer from the entropy mode to the adiabatic mode ($\Theta \neq 0$). In a multi-field DBI inflation model, the tensor-to-scalar ratio is given by
\begin{equation}
 r \equiv \frac{P_{\mathcal{T}}}{P_{\mathcal{R}}} = \left. 16 \epsilon c_{s} \right\rvert_{*} \cos^{2} {\Theta},
\label{tensortoscalar}
\end{equation}
from Eq. (\ref{chaptertwotransferdefined}) and the amplitude of the tensor perturbation
\begin{equation}
 P_{\mathcal{T}} = \left. \frac{2 H^{2}}{\pi^{2}} \right\rvert_{*}. 
\end{equation}
In single field cases, we have $\Theta = 0$. It is clear that the more conversion of the entropy perturbation we have, the smaller the value of $r$ becomes. Therefore, the lower bound (\ref{lower}) no longer exists in multi-field DBI inflation models. 

\subsection{Observational constraints combined with the analytic formulae}\label{observationalconstraintlastsubsec}
\begin{figure}
  \centering
   \begin{tabular}{cc} 
    \includegraphics[width=70mm]{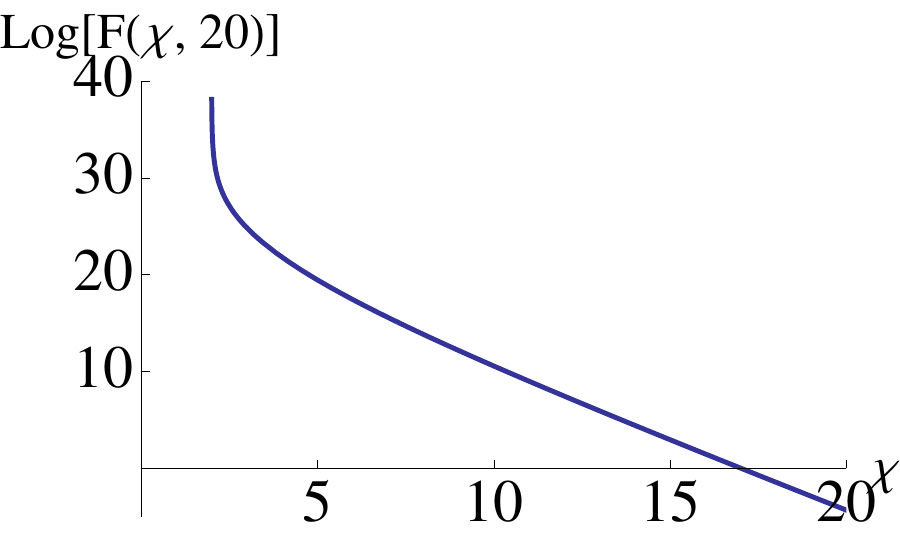}&
    \includegraphics[width=70mm]{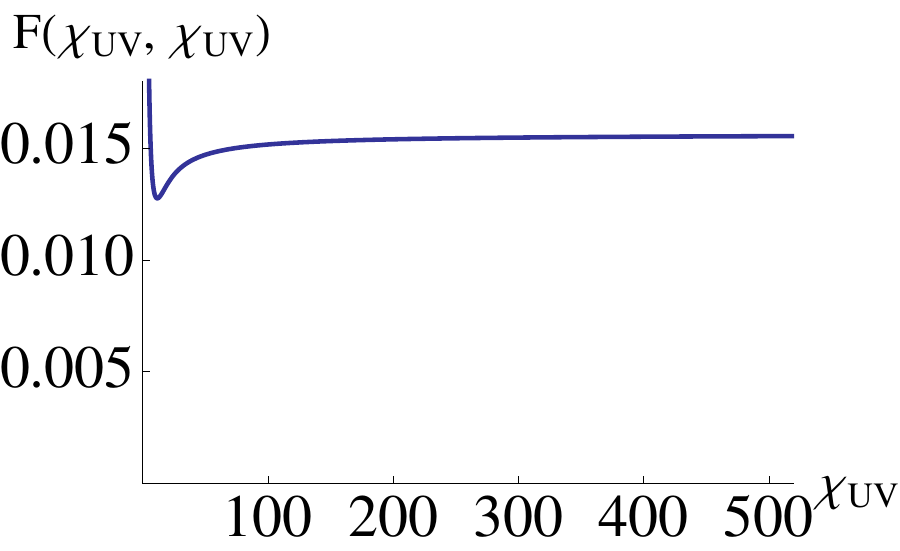}
  \end{tabular}
  \caption[Geometrical function]{Left: semi-log plot of $F\left(\chi, \chi_{\rm{UV}}\right)$ with respect to $\chi$ for $\chi_{\rm{UV}} = 20$. It keeps decreasing exponentially. Right: plot of $F\left(\chi_{\rm{UV}}, \chi_{\rm{UV}} \right)$ with respect to $\chi_{\rm{UV}}$. It takes a constant value asymptotically. }\label{geometricalfunctionplots}
\end{figure}

As shown in subsection \ref{sebsec:microphysical}, the microphysical constraint that disfavours the single field DBI inflation models is possibly satisfied when the power spectrum of the curvature perturbation is enhanced after the horizon exit. In this section, we consider the cases where the dynamics is effectively single field until the perturbations considered are stretched to super-horizon scales as in the example in subsection \ref{subsec:approximationofspinflation}. The enhancement is quantified by the transfer function as in Eq. (\ref{chaptertwotransferdefined}). Using Eq. (\ref{chaptertwotransferdefined}), the ratio of the power spectrum of the curvature perturbation at the end of inflation $\mathcal{P}_{\mathcal{R}}$ to the power spectrum of the curvature perturbation around horizon crossing $\mathcal{P}_{\mathcal{R}_{*}}$ is given by
\begin{equation}\label{spincosthetadefinitionfinal}
\cos^{-2}{\Theta} = \frac{\mathcal{P}_{\mathcal{R}}}{\mathcal{P}_{\mathcal{R}_{*}}}.
\end{equation}
Because $\mathcal{P}_{\mathcal{R}}$ at the end of inflation needs to satisfy the constraint by the Planck satellite observations \cite{Ade:2013b}, we have $\mathcal{P}_{\mathcal{R}} \sim 2.2 \times 10^{-9}$. Because we need $\cos^{-2}{\Theta} \gg 1$ to make the multi-field DBI inflation model compatible with the Planck satellite observations for the equilateral non-Gaussianity, we require
\begin{equation}\label{spinanaconpr}
\mathcal{P}_{\mathcal{R}_{*}} < 10^{-9}.
\end{equation}
Using the approximated analytic expression (\ref{curvaturespectrumexampledbisingle}), Eq. (\ref{spinanaconpr}) gives the lower bound of $\lambda$ as
\begin{equation}\label{lowerboundlambdaspin}
\lambda > \frac{10^{9}}{4 \pi^2 \epsilon^4},
\end{equation}
with the slow-roll parameter $\epsilon$. From Eqs. (\ref{dbimultiequilateralfinalresult}) and (\ref{spincosthetadefinitionfinal}), we obtain
\begin{equation}\label{fnlequilspinfirst}
\begin{split}
f_{\rm{NL}}^{equil} &\approx -\frac{\cos^{2}{\Theta}}{3 c_{s}^{2}}\\
&= -\frac{1}{3 c_{s}^{2}}\frac{\mathcal{P}_{\mathcal{R}_{*}}}{\mathcal{P}_{\mathcal{R}}}. 
\end{split}
\end{equation}
Using the constraint on $f_{\rm{NL}}^{equil}$ by the Planck satellite observations $\left\lvert f_{\rm{NL}}^{equil} \right\rvert < 100$, Eq. (\ref{fnlequilspinfirst}) leads to
\begin{equation}\label{spinconstraintprandcs}
\frac{\mathcal{P}_{\mathcal{R}_{*}}}{c_{s}^{2}} < 6 \times 10^{-7},
\end{equation}
where we have used $\mathcal{P}_{\mathcal{R}} \sim 2.2 \times 10^{-9}$. The inequality (\ref{spinconstraintprandcs}) is rewritten as
\begin{equation}\label{constraintspinlastchapphi}
\frac{10^{7}}{6 \pi^{2} \epsilon^{6} \lambda} < \left(\frac{\phi}{M_{\rm{P}}} \right)^{4},
\end{equation}
using Eq. (\ref{curvaturespectrumexampledbisingle}) and the relation
\begin{equation}
c_{s} = \frac{\epsilon}{2} \left(\frac{\phi}{M_{\rm{P}}} \right)^{2},
\end{equation}
which is derived from Eq. (\ref{alltheresultsindbisingleexample}). From Eqs. (\ref{massunitdefinitionspin}) and (\ref{spinexamplecanonicalfield}), the canonical field in the Planck units is given by
\begin{equation}\label{phioverplanckmassspinconstraint}
\begin{split}
\frac{\phi \left(\chi\right)}{M_{\rm{P}}} &= \frac{\sqrt{T_{3}} \kappa^{2/3}}{M_{\rm{P}}} \frac{1}{\sqrt{6}}\int^{\chi}_{0}\frac{dx}{K\left(x\right)}\\
&= \sqrt{\bar{T}_{3}} \bar{\kappa}^{2/3} \frac{\mathcal{M}}{M_{\rm{P}}} \frac{1}{\sqrt{6}}\int^{\chi}_{0}\frac{dx}{K\left(x\right)}\\
&= \sqrt{\frac{6 \pi}{\bar{\kappa}^{4/3}g_{\rm{s}}M^{2}\bar{T}_{3} N J \left(\chi_{\rm{UV}}\right)}} \sqrt{\bar{T}_{3}} \bar{\kappa}^{2/3} \frac{1}{\sqrt{6}}\int^{\chi}_{0}\frac{dx}{K\left(x\right)}\\
&= \sqrt{\frac{\pi}{g_{\rm{s}}M^{2} N J \left(\chi_{\rm{UV}}\right)}} \int^{\chi}_{0}\frac{dx}{K\left(x\right)}.
\end{split}
\end{equation}
From Eqs. (\ref{constraintspinlastchapphi}) and (\ref{phioverplanckmassspinconstraint}), we obtain the inequality
\begin{equation}
\left(\sqrt{\frac{\pi}{g_{\rm{s}}M^{2} N J \left(\chi_{\rm{UV}}\right)}} \int^{\chi}_{0}\frac{dx}{K\left(x\right)} \right)^{4} > \frac{10^{7}}{6 \pi^{2} \epsilon^{6} \lambda},
\end{equation}
which leads to
\begin{equation}\label{preupperboundforlambdaspin}
\frac{\pi^2}{N^{2} J \left(\chi_{\rm{UV}}\right)^{2}} \left[ \frac{27}{64 \pi^3 \lambda} \left(\ln{\frac{r\left(\chi \right)^{3}}{\kappa^2}} + \ln{\frac{4\sqrt{2}}{3\sqrt{3}}} - \frac{1}{4} \right) \right]^{2} \left(\int^{\chi}_{0}\frac{dx}{K\left(x\right)} \right)^{4} > \frac{10^{7}}{6 \pi^{2} \epsilon^{6} \lambda}. 
\end{equation}
from Eq. (\ref{largechilambda}). Simplifying Eq. (\ref{preupperboundforlambdaspin}), we obtain the upper bound of $\lambda$ as
\begin{equation}\label{lambdaupperboundspin}
\lambda < \frac{3}{2 \pi^{2}} \left(\frac{27}{32}\right)^{2} 10^{-7} \epsilon^{6} \frac{\left(\ln{\frac{r\left(\chi \right)^{3}}{\kappa^2}} + \ln{\frac{4\sqrt{2}}{3\sqrt{3}}} - \frac{1}{4} \right)^{2}}{J \left(\chi_{\rm{UV}}\right)^{2}} \left(\int^{\chi}_{0}\frac{dx}{K\left(x\right)} \right)^{4}.
\end{equation}
Because we have both the lower bound (\ref{lowerboundlambdaspin}) and the upper bound (\ref{lambdaupperboundspin}) of $\lambda$, the lower bound must be smaller than the upper bound
\begin{equation}
\frac{10^{9}}{4 \pi^2 \epsilon^4} < \frac{3}{2 \pi^{2}} \left(\frac{27}{32}\right)^{2} 10^{-7} \epsilon^{6} \frac{\left(\ln{\frac{r\left(\chi \right)^{3}}{\kappa^2}} + \ln{\frac{4\sqrt{2}}{3\sqrt{3}}} - \frac{1}{4} \right)^{2}}{J \left(\chi_{\rm{UV}}\right)^{2}} \left(\int^{\chi}_{0}\frac{dx}{K\left(x\right)} \right)^{4},
\end{equation}
which is rewritten as
\begin{equation}\label{fchiconditionfinalspin}
F\left(\chi, \chi_{\rm{UV}} \right) < 4.27 \times 10^{-16} \epsilon^{10},
\end{equation}
with
\begin{equation}\label{explicitdefinitionoffchichiuvspin}
F\left(\chi, \chi_{\rm{UV}} \right) \equiv \frac{J \left(\chi_{\rm{UV}}\right)^{2}}{\left[\ln{ \left(\frac{1}{\sqrt{6}}\int^{\chi}_{0}\frac{dx}{K\left(x\right)} \right)^3} + \ln{\frac{4\sqrt{2}}{3\sqrt{3}}} - \frac{1}{4} \right]^{2} \left(\int^{\chi}_{0}\frac{dx}{K\left(x\right)} \right)^{4}},
\end{equation}
where we have used Eq. (\ref{eqforproperradial}). If the condition (\ref{fchiconditionfinalspin}) is not satisfied, $\lambda$ cannot take any value that is larger than the lower bound (\ref{lowerboundlambdaspin}) and smaller than the upper bound (\ref{lambdaupperboundspin}) at the same time. Because the function $F\left(\chi \right)$ is dependent only on $\chi$ and $\chi_{UV}$, this is a general condition that is independent of all other parameters. Numerically, we obtain
\begin{equation}
L \left(\chi_{N} \right) \equiv \ln{ \left(\frac{1}{\sqrt{6}}\int^{\chi_{N}}_{0}\frac{dx}{K\left(x\right)} \right)^3} + \ln{\frac{4\sqrt{2}}{3\sqrt{3}}} - \frac{1}{4} = 0,
\end{equation}
where $\chi_{N} = 1.9966$. As $\chi$ increases from $\chi_{N}$, $L \left(\chi \right)$ increases monotonically because we have
\begin{equation}
\frac{d}{d \chi} \left(\int^{\chi}_{0}\frac{dx}{K\left(x\right)} \right) = \frac{1}{K\left(x\right)} >0.
\end{equation}
Because we consider the case $\chi \gg 1$, we study the behaviour of $F\left(\chi, \chi_{\rm{UV}} \right)$ only in the region $\chi > 2$ below. Therefore, the denominator of $F\left(\chi, \chi_{\rm{UV}} \right)$ in Eq. (\ref{explicitdefinitionoffchichiuvspin}) is a monotonically increasing function with $\chi$. This means that $\chi=\chi_{\rm{UV}}$ minimises $F\left(\chi, \chi_{\rm{UV}} \right)$ and the condition (\ref{fchiconditionfinalspin}) is rewritten as
\begin{equation}\label{fchiuvfinalconditionspin}
F\left(\chi_{\rm{UV}}, \chi_{\rm{UV}} \right) < 4.27 \times 10^{-16} \epsilon^{10} < 4.27 \times 10^{-16},
\end{equation}
where we used $\epsilon < 1$ during inflation. Choosing $\chi = \chi_{\rm{UV}}$ means considering the perturbation that exits the horizon when the brane is at $\chi = \chi_{\rm{UV}}$. As shown in the left panel of figure \ref{geometricalfunctionplots}, $F\left(\chi, \chi_{\rm{UV}} \right)$ keeps decreasing exponentially. The plot is for $\chi_{\rm{UV}} = 20$ and $F\left(\chi_{\rm{UV}}, \chi_{\rm{UV}} \right) \approx 0.0134$ in this case. This does not satisfy the condition (\ref{fchiuvfinalconditionspin}). 

Let us show the behaviour of $F\left(\chi_{\rm{UV}}, \chi_{\rm{UV}} \right)$ with respect to $\chi_{\rm{UV}}$ below. For large $\chi$, we obtain
\begin{equation}
K\left(\chi \right) \approx 2^{1/3} \exp{\left(-\frac{1}{3} \chi\right)},
\end{equation}
which leads to
\begin{equation}\label{spincoordinateapproximation}
\int^{\chi}_{0}\frac{dx}{K\left(x\right)} \approx \frac{3}{2^{1/3}} \exp{\left(\frac{1}{3} \chi \right)}. 
\end{equation}
Using Eq. (\ref{spincoordinateapproximation}), we obtain
\begin{equation}\label{spinapproximationofthelnfunctioninf}
\ln{ \left(\frac{1}{\sqrt{6}}\int^{\chi}_{0}\frac{dx}{K\left(x\right)} \right)^3} + \ln{\frac{4\sqrt{2}}{3\sqrt{3}}} - \frac{1}{4} \approx \chi. 
\end{equation}
The function $I\left( \chi \right)$ in Eq. (\ref{eqforfunctioni}) is approximated as
\begin{equation}\label{chapfourapproxifuninlargechi}
I \left( \chi \right) \approx \frac{3}{4^{1/3}} \chi \exp{\left(- \frac{4}{3} \chi \right)},
\end{equation}
for large $\chi$ from the approximated expression in \cite{spinflation}. For sufficiently large $\chi_{\rm{UV}}$, we obtain
\begin{equation}\label{spinfinalapproxforjuv}
\begin{split}
J \left(\chi_{\rm{UV}} \right) &= \int^{\chi_{\rm{UV}}}_{0} d\chi I \left(\chi \right) \sinh^{2}{\chi}\\
&\approx \int^{\chi_{\rm{UV}}}_{\chi_{\rm{t}}} d\chi I \left(\chi \right) \sinh^{2}{\chi}\\
&\approx \frac{3^{2}}{2^{11/3}} \chi_{\rm{UV}} \exp{\left(\frac{2}{3} \chi_{\rm{UV}} \right)},
\end{split}
\end{equation}
using Eq. (\ref{chapfourapproxifuninlargechi}) where $1 \ll \chi_{\rm{t}} \ll \chi_{\rm{UV}}$. From Eqs. (\ref{explicitdefinitionoffchichiuvspin}), (\ref{spincoordinateapproximation}), (\ref{spinapproximationofthelnfunctioninf}) and (\ref{spinfinalapproxforjuv}), we obtain
\begin{equation}
F\left(\chi_{\rm{UV}}, \chi_{\rm{UV}} \right) \approx \frac{1}{2^{6}} \approx 0.0156,
\end{equation}
for large $\chi_{\rm{UV}}$. In the right panel of figure \ref{geometricalfunctionplots}, we see that $F\left(\chi_{\rm{UV}}, \chi_{\rm{UV}} \right)$ actually approaches $0.0156$. It also shows that $F\left(\chi_{\rm{UV}}, \chi_{\rm{UV}} \right)$ does not become smaller than $0.01$ in the region $2 <\chi_{\rm{UV}}$ before it becomes constant. Therefore, we conclude that the necessary condition (\ref{fchiuvfinalconditionspin}) is not satisfied regardless of the value of $\chi_{\rm{UV}}$. Because the condition (\ref{fchiuvfinalconditionspin}) is independent of any other parameter, this model is excluded by the observations in the regime where we can use those approximated formulae. Note that this strong constraint comes from the fact that the sound speed and the amplitude of the curvature perturbation is controlled essentially by one factor that consists of the model parameters $g_{\rm{s}} M^2$ as in Eqs. (\ref{largechilambda}) and (\ref{phioverplanckmassspinconstraint}). Due to this relation, it is not possible to satisfy (\ref{spinanaconpr}) and (\ref{spinconstraintprandcs}) simultaneously.

\section{Summary and Discussions}
\label{summarydiscussions}
In this paper, we studied the DBI inflation model with the simplest two-field potential derived in string theory \cite{spinflation}. When we consider the cases where we have the effective single field dynamics around horizon crossing, the model is approximated with a simple model studied in \cite{Alishahiha:2004}. After the horizon exit, the power spectrum of the curvature perturbation is significantly enhanced if the trajectory of the brane in the field space is bent sharply. Because all the analytic expressions derived in section \ref{sec:nongaussianity} are valid with the effective single field dynamics around the horizon crossing, we can predict the value of the non-Gaussianity parameters. It has been shown that the model is excluded with the constraints on the power spectrum of the curvature perturbation and $f_{\rm{NL}}^{equil}$ regardless of the model parameters when we take into account the constraint on the volume of the internal space \cite{Baumann:2007}. 

We need further analysis if the coupling between the adiabatic and entropy perturbations is not negligible around horizon crossing as $\xi/aH > 1$. In such cases, the adiabatic perturbation is coupled to the entropy perturbation around horizon crossing and the power spectrum of the curvature perturbation around horizon crossing can no longer be estimated with Eq. (\ref{curvaturepowerspectrumhorizon}). The expression for $f_{\rm{NL}}^{equil}$ (\ref{dbimultiequilateralfinalresult}) is also not valid in such cases because we used the expression for the curvature power spectrum which is valid only with a small coupling around horizon crossing  in deriving this expression. Because the analytic formulae in subsection \ref{subsec:approximationofspinflation} are derived assuming the single field dynamics, the conclusion in subsection \ref{observationalconstraintlastsubsec} is no longer valid in cases with large couplings around horizon crossing. However, our numerical calculations show that it is difficult to maintain the almost scale-invariant curvature power spectrum which is compatible with the observations when the coupling is large around horizon crossing. 

If the coupling around horizon crossing is large, we need to calculate the non-Gaussianity performing the full calculations with the in-in formalism. In \cite{Chen:2010b}, the authors have performed similar calculations in quasi-single field inflation where the coupling is not negligible around horizon crossing. They studied a model where there is one slow-roll direction while all other isocurvature fields have masses at least of the order of $H$ and obtained large bispectra whose shape is between the equilateral and local shapes. It would be interesting to apply their method to study the shape of non-Gaussianity in the case with the large coupling to know whether the stringent microphysical constraint can be avoided. 

Though we studied the simplest potential that takes into account the leading order correction to the potential in this paper, the shape of the potential can be more complicated depending on the embbeding of branes in the internal space. For example, in \cite{Chen:2010}, they obtained a potential where a waterfall phase transition connects two different radial trajectories. In \cite{Kidani:2012}, for the first time, we quantified the effect of the angular dynamics on observables using a toy model representing this type of the potential. We demonstrated that all the Planck observational constraints can be satisfied, except for the constraint on $f_{\rm{NL}}^{local}$, while obeying the bound on the tensor-to-scalar ratio imposed in string theory models. In general, the large conversion creates large local type non-Gaussianity. In our model, this is indeed the case and we expect that large equilateral non-Gaussianity is generally accompanied by large local type non-Gaussainity in multi-field DBI models. Those studies show that the precise measurement of the CMB anisotropies makes it possible to test DBI inflation models effectively once concrete potentials are given. Therefore, it is important to study further the realisation of DBI inflation models in string theory. 

While we are completing this paper, a new exciting measurement of the B-mode polarisation was made by the BICEP2 experiment suggesting $r = 0.20^{+0.07}_{-0.05}$ at $7.0 \sigma$ \cite{Ade:2014xna}. Even though the result by BICEP2 should be confirmed with measurements of the CMB polarisation in other experiments such as the PLANCK satellite observation, $r \simeq 0.2$ would not satisfy the microphysical constraint (\ref{rcondition}) for the single field DBI inflation. There are several ways of relaxing the microphysical constraints using wrapped branes \cite{Becker:2007, McAllister:2008, Avgoustidis:2009}, multiple branes \cite{Avgoustidis:2008zu, Krause:2008, Ashoorioon:2009, Ashoorioon:2011, Ashoorioon:2014} or multiple fields as explained in subsection \ref{sebsec:microphysical}. With those ideas, DBI inflation models could be compatible with the BICEP2 result, however, the DBI inflation models will face a significant challenge. This can be seen from the formula for the tensor-to-scalar ratio $r$, (\ref{tensortoscalar}). Both the small sound speed $c_s <1$ and the multi-field effect $\cos^2 \Theta <1$ suppress the tensor-to-scalar ratio. Although it is still possible to find a model that produces $r=0.2$ in DBI inflation models \cite{Avgoustidis:2008zu}, it requires more elaborated constructions. 
 
\begin{acknowledgments}
We would like to thank Jon Emery and Gianmassimo Tasinato for useful discussions. We also would like to thank Ruth Gregory and Dariush Kaviani for providing us with their numerical code and for helping us to modify the code in order to obtain the numerical results. T. K. and K. K. were supported by the Leverhulme trust. K. K. is supported by the UK Science and Technology Facilities Council grants number ST/K00090/1 and number ST/L005573/1. 

\end{acknowledgments}


\begin{thebibliography}{99}

\bibitem{Komatsu:2010}
  E.~Komatsu, {\it et al.}  [WMAP Collaboration],
  Astrophys.\ J.\ Suppl.\  {\bf 192}, 18 (2011)
 [\href{http://arxiv.org/abs/1001.4538}{{\tt arXiv:1001.4538 [astro-ph.CO]}}].

\bibitem{Kaiser:2013}
  D. I. Kaiser, E. A. Mazenc and E. I. Sfakianakis,
  Phys. Rev. D {\bf 87} (2013) 064004
 [\href{http://arxiv.org/abs/1210.7487}{{\tt arXiv:1210.7487 [astro-ph.CO]}}].

\bibitem{Kaiser:2014}
  D. I. Kaiser and E. I. Sfakianakis,
  Phys. Rev. Lett. {\bf 112} (2014) 011302
 [\href{http://arxiv.org/abs/1304.0363}{{\tt arXiv:1304.0363 [astro-ph.CO]}}].

\bibitem{Schutz:2014}
  K. Schutz, E. I. Sfakianakis and D. I. Kaiser,
  Phys. Rev. D {\bf 89} (2014) 064044
 [\href{http://arxiv.org/abs/1310.8285}{{\tt arXiv:1310.8285 [astro-ph.CO]}}].

\bibitem{Naruko:2009}
  A. Naruko and M. Sasaki,
  Prog. Theor. Phys. {\bf121} (2009) 193-210
 [\href{http://arxiv.org/abs/0807.0180}{{\tt arXiv:0807.0180 [astro-ph]}}].

\bibitem{Ade:2013}
Planck Collaboration: P. A. R. Ade, N. Aghanim, C. Armitage-Caplan, {\it et al.},
 [\href{http://arxiv.org/abs/1303.5082}{{\tt arXiv:1303.5082 [astro-ph.CO]}}].

\bibitem{Silverstein:2004}
  E.~Silverstein and D.~Tong,
  Phys. Rev. D {\bf 70} (2004) 103505
 [\href{http://arxiv.org/abs/hep-th/0310221}{{\tt arXiv:hep-th/0310221}}].

\bibitem{Alishahiha:2004}
  M.~Alishahiha , E.~Silverstein and D.~Tong,
  Phys. Rev. D {\bf 70} (2004) 123505
[\href{http://arxiv.org/abs/hep-th/0404084}{{\tt arXiv:hep-th/0404084}}].

\bibitem{Underwood:2008}
  B.~Underwood,
  Phys. Rev. D {\bf 78} (2008) 023509
 [\href{http://arxiv.org/abs/0802.2117}{{\tt arXiv:0802.2117 [hep-th]}}].

\bibitem{Chen:2006nt}
  X.~Chen, M.~-x.~Huang, S.~Kachru and G.~Shiu,
JCAP {\bf 0701}, 002 (2007) 
[\href{http://arxiv.org/abs/hep-th/0605045}{{\tt arXiv:hep-th/0605045}}].

\bibitem{Baumann:2007}
  D.~Baumann and L.~McAllister,
  Phys. Rev. D {\bf 75} (2007) 123508
 [\href{http://arxiv.org/abs/hep-th/0610285}{{\tt arXiv:hep-th/0610285}}].

\bibitem{Lidsey:2007}
  J.~Lidsey and I.~Huston,
  JCAP {\bf 0707} (2007) 002
 [\href{http://arxiv.org/abs/0705.0240}{{\tt arXiv:0705.0240 [hep-th]}}].

\bibitem{Ian:2008}
  I.~Huston, J.~E.~Lidsey, S.~Thomas and J.~Ward,
  JCAP {\bf 0805} (2008) 016
 [\href{http://arxiv.org/abs/0708.4321}{{\tt arXiv:0708.4321 [hep-th]}}].

\bibitem{Kobayashi:2007hm}
  T.~Kobayashi, S.~Mukohyama and S.~Kinoshita,
JCAP {\bf 0801}, 028 (2008) 
[\href{http://arxiv.org/abs/0708.4285}{{\tt arXiv:0708.4285 [hep-th]}}].

\bibitem{Bean:2008}
  R.~Bean, X.~Chen, H.~Peiris and J.~Xu,
  Phys. Rev. D {\bf 77} (2008) 023527
[\href{http://arxiv.org/abs/0710.1812}{{\tt arXiv:0710.1812 [hep-th]}}].

\bibitem{Gordon:2001}
  C.~Gordon, D.~Wands, B.~A.~Bassett and R.~Maartens,
  Phys. Rev. D {\bf 63} (2001) 023506
 [\href{http://arxiv.org/abs/astro-ph/0009131}{{\tt arXiv:astro-ph/0009131}}].

\bibitem{Langlois:2008wt}
  D.~Langlois, S.~Renaux-Petel, D.~A.~Steer and T.~Tanaka,
Phys.\ Rev.\ Lett.\  {\bf 101} (2008) 061301
[\href{http://arxiv.org/abs/0804.3139}{{\tt arXiv:0804.3139 [hep-th]}}].  

\bibitem{Langlois:2008qf}
  D.~Langlois, S.~Renaux-Petel, D.~A.~Steer and T.~Tanaka,
Phys.\ Rev.\ D {\bf 78}, 063523 (2008) 
[\href{http://arxiv.org/abs/0806.0336}{{\tt arXiv:0806.0336 [hep-th]}}].  

\bibitem{Langlois:2009}
  D.~Langlois, S.~Renaux-Petel and D.~A.~Steer,
  JCAP {\bf 0904} (2009) 021
 [\href{http://arxiv.org/abs/0902.2941}{{\tt arXiv:0902.2941 [hep-th]}}].

\bibitem{Arroja:2008yy}
  F.~Arroja, S.~Mizuno and K.~Koyama,
  JCAP {\bf 0808}, 015 (2008) 
  [\href{http://arxiv.org/abs/0806.0619}{{\tt arXiv:0806.0619 [astro-ph]}}].  
  
\bibitem{Kidani:2012}
  T. Kidani, K. Koyama and S. Mizuno,
Phys.\ Rev. \ D {\bf 86}, 083503 (2012) 
[\href{http://arxiv.org/abs/1207.4410}{{\tt arXiv:1207.4410 [astro-ph.CO]}}].

\bibitem{Baumann:2007ah}
  D.~Baumann, A.~Dymarsky, I.~R.~Klebanov and L.~McAllister,
  JCAP {\bf 0801}, 024 (2008) 
  [\href{http://arxiv.org/abs/0706.0360}{{\tt arXiv:0706.0360 [hep-th]}}].  
  
\bibitem{spinflationbefore}
  D.~A.~Easson, R.~Gregory, D.~F.~Mota, G.~Tasinato and I.~Zavala,
JCAP {\bf 0802}, 010 (2008) 
[\href{http://arxiv.org/abs/0709.2666}{{\tt arXiv:0709.2666 [hep-th]}}].
 
 \bibitem{spinflation}
 R.~Gregory and D.~Kaviani,
  JHEP {\bf 1201}, 037 (2012)
 [\href{http://arxiv.org/abs/1107.5522}{{\tt arXiv:1107.5522 [hep-th]}}].

\bibitem{local}
  F.~Vernizzi and D.~Wands,
  JCAP {\bf 0605} (2006) 019
  [\href{http://arxiv.org/abs/astro-ph/0603799}{{\tt arXiv:astro-ph/0603799}}];
  S.~Yokoyama, T.~Suyama and T.~Tanaka,
  JCAP {\bf 0707} (2007) 013
  [\href{http://arxiv.org/abs/0705.3178}{{\tt arXiv:0705.3178 [astro-ph]}}];
  C.~T.~Byrnes, K.~-Y.~Choi and L.~M.~H.~Hall,
  JCAP {\bf 0810}, 008 (2008) 
   [\href{http://arxiv.org/abs/0807.1101}{{\tt arXiv:0807.1101 [astro-ph]}}]; 
  C.~T.~Byrnes, K.~-Y.~Choi and L.~M.~H.~Hall,
  JCAP {\bf 0902}, 017 (2009) 
   [\href{http://arxiv.org/abs/0812.0807}{{\tt arXiv:0812.0807 [astro-ph]}}]; 
  C.~T.~Byrnes and G.~Tasinato,
  JCAP {\bf 0908}, 016 (2009) 
   [\href{http://arxiv.org/abs/0906.0767}{{\tt arXiv:0906.0767 [astro-ph.CO]}}]; 
  M. ~Sasaki,
  Prog. Theor. Phys. {\bf 120} 159-174 (2008)
 [\href{http://arxiv.org/abs/0805.0974}{{\tt arXiv:0805.0974 [astro-ph]}}];
  J.~Meyers and N.~Sivanandam,
  Phys. Rev. D {\bf 83} (2011) 103517
  [\href{http://arxiv.org/abs/1011.4934}{{\tt arXiv:1011.4934 [astro-ph.CO]}}];
  J.~Frazer and A.~R.~Liddle,
  JCAP {\bf 1202} (2012) 039
  [\href{http://arxiv.org/abs/1111.6646}{{\tt arXiv:1111.6646 [astro-ph.CO]}}];
  K.~Koyama, S.~Mizuno, F.~Vernizzi and D.~Wands,
  JCAP {\bf 0711} (2007) 024
 [\href{http://arxiv.org/abs/0708.4321}{{\tt arXiv:0708.4321 [hep-th]}}].;
  G.~I.~Rigopoulos, E.~P.~S.~Shellard and B.~J.~W.~van Tent,
  Phys. Rev. D {\bf 73} (2006) 083522
 [\href{http://arxiv.org/abs/astro-ph/0506704}{{\tt arXiv:astro-ph/0506704}}];
  G.~I.~Rigopoulos, E.~P.~S.~Shellard and B.~J.~W.~van Tent,
  Phys. Rev. D {\bf 76} (2007) 083512
  [\href{http://arxiv.org/abs/astro-ph/0511041}{{\tt arXiv:astro-ph/0511041}}]; 
  A.~Mazumdar and L.~Wang,
  [\href{http://arxiv.org/abs/1203.3558}{{\tt arXiv:1203.3558 [astro-ph.CO]}}]. 

\bibitem{Wands:2010af}
  D.~Wands,
  Class.\ Quant.\ Grav.\  {\bf 27}, 124002 (2010)  
     [\href{http://arxiv.org/abs/1004.0818}{{\tt arXiv:1004.0818 [astro-ph.CO]}}]. 
     
\bibitem{Koyama:2010}
  K.~Koyama,
  Class. Quantum Grav. {\bf 27} (2010) 124001
   [\href{http://arxiv.org/abs/1002.0600}{{\tt arXiv:1002.0600 [hep-th]}}]. 

\bibitem{Babich:2004}
  D.~Babich, P.~Creminelli and M.~Zaldarriaga,
  JCAP {\bf 0408} (2004) 009
  [\href{http://arxiv.org/abs/astro-ph/0405356}{{\tt arXiv:astro-ph/0405356}}]. 

\bibitem{Creminelli:2006}
  P. Creminelli, A. Nicolis, L. Senatore, M. Tegmark and M. Zaldarriaga,
  JCAP {\bf 0605} (2006) 004
  [\href{http://arxiv.org/abs/astro-ph/0509029}{{\tt arXiv:astro-ph/0509029}}]. 

\bibitem{RenauxPetel:2009sj}
  S.~Renaux-Petel,
  JCAP {\bf 0910} (2009) 012 
  [\href{http://arxiv.org/abs/0907.2476}{{\tt arXiv:0907.2476 [hep-th]}}]. 

\bibitem{Klebanov:2000}
  I. R. Klebanov  and A. A. Tseytlin,
  JHEP 0008 (2000) 052
  [\href{http://arxiv.org/abs/hep-th/0007191}{{\tt arXiv:hep-th/0007191}}]. 
 
\bibitem{Baumann:2010}
D. Baumann, A. Dymarsky, S. Kachru, I. R. Klebanov and L. McAllister, 
 (2009)
  [\href{http://arxiv.org/abs/1001.5028}{{\tt arXiv:1001.5028 [hep-th]}}].

\bibitem{Thanks:Ruth}
We would like to thank R. Gregory for providing us with their Mathematica code and confirming the numerical issue in their original results presented in \cite{spinflation}.

\bibitem{Langlois:2008b}
 D. Langlois and S. Renaux-Petel,
  JCAP {\bf 0804} (2008) 017
  [\href{http://arxiv.org/abs/0801.1085}{{\tt arXiv:0801.1085 [hep-th]}}].

\bibitem{Kodama:1984}
H. Kodama and M. Sasaki,
  Prog.Theor.Phys.Suppl. {\bf 78} (1984) 001

\bibitem{Nalson:2013}
 E. Nalson, A. J. Christopherson, I. Huston and K. A. Malik,
  Class. Quantum Grav. {\bf 30} (2013) 065008
[\href{http://arxiv.org/abs/1111.6940}{{\tt arXiv:1111.6940 [astro-ph.CO]}}].

\bibitem{Nalson:2013b}
 E. Nalson, A. J. Christopherson, I. Huston and K. A. Malik,
  (2013)
[\href{http://arxiv.org/abs/1301.7613}{{\tt arXiv:1301.7613 [astro-ph.CO]}}].

\bibitem{Maldacena:2005}
 J. Maldacena,
  JHEP {\bf 0305} (2003) 013
[\href{http://arxiv.org/abs/astro-ph/0210603}{{\tt arXiv:astro-ph/0210603}}].

\bibitem{Chen:2007}
 X. Chen, M. Huang, S. Kachru and G. Shiu,
  JCAP {\bf 0701} (2007) 002
[\href{http://arxiv.org/abs/hep-th/0605045}{{\tt arXiv:hep-th/0605045}}].

\bibitem{Weinberg:2005}
S. Weinberg,
 Phys. Rev. {\bf D. 72} (2005) 043514
[\href{http://arxiv.org/abs/hep-th/0506236}{{\tt arXiv:hep-th/0506236}}].

\bibitem{Lyth:2005}
L. Boubekeur and D. H. Lyth,
 JCAP {\bf 0507} (2005) 010
[\href{http://arxiv.org/abs/hep-ph/0502047}{{\tt arXiv:hep-ph/0502047}}].

\bibitem{Ade:2013c}
Planck Collaboration: P. A. R. Ade, N. Aghanim, C. Armitage-Caplan, {\it et al.},
 (2013)
[\href{http://arxiv.org/abs/1303.5084}{{\tt arXiv:1303.5084 [astro-ph.CO]}}].

\bibitem{Ade:2013b}
Planck Collaboration: P. A. R. Ade, N. Aghanim, C. Armitage-Caplan, {\it et al.},
 (2013)
[\href{http://arxiv.org/abs/1303.5076}{{\tt arXiv:1303.5076 [astro-ph.CO]}}].

\bibitem{Chen:2010b}
X. Chen, Y. Wang, 
 JCAP {\bf 1004} (2010) 027,
  [\href{http://arxiv.org/abs/0911.3380}{{\tt arXiv:0911.3380 [hep-th]}}].
  
\bibitem{Chen:2010}
H. Chen, J. Gong, K. Koyama and G. Tasinato,
   JCAP {\bf 1011} (2010) 034,
  [\href{http://arxiv.org/abs/1007.2068}{{\tt arXiv:1007.2068 [hep-th]}}].
  

\bibitem{Ade:2014xna} 
  P.~A.~R.~Ade {\it et al.}  [BICEP2 Collaboration],
  arXiv:1403.3985 [astro-ph.CO].
  
  \bibitem{Becker:2007}
M. Becker, L. Leblond and S. Shandera,
  Phys. Rev. {\bf D. 76} (2007) 123516
  [\href{http://arxiv.org/abs/0709.1170}{{\tt arXiv:0709.1170 [hep-th]}}].

\bibitem{McAllister:2008}
L. McAllister, E. Silverstein and A. Westphal,
  [\href{http://arxiv.org/abs/0808.0706}{{\tt arXiv:0808.0706 [hep-th]}}].
    
\bibitem{Avgoustidis:2009}
A. Avgoustidis and I. Zavala,
   JCAP {\bf 0901} (2009) 045,
  [\href{http://arxiv.org/abs/0810.5001}{{\tt arXiv:0810.5001 [hep-th]}}].
  
  \bibitem{Avgoustidis:2008zu} 
    A.~Avgoustidis and I.~Zavala,
    JCAP {\bf 0901}, 045 (2009)
    [arXiv:0810.5001 [hep-th]].
  
  \bibitem{Krause:2008}
A. Krause,
   JCAP {\bf 0807} (2008) 001,
  [\href{http://arxiv.org/abs/0708.4414}{{\tt arXiv:0708.4414 [hep-th]}}].

  \bibitem{Ashoorioon:2009}
A. Ashoorioon, H. Firouzjahi and M. M. Sheikh-Jabbari,
   JCAP {\bf 0906} (2009) 018,
  [\href{http://arxiv.org/abs/0903.1481}{{\tt arXiv:0903.1481 [hep-th]}}].

  \bibitem{Ashoorioon:2011}
A. Ashoorioon and M. M. Sheikh-Jabbari,
   JCAP {\bf 1106} (2011) 014,
  [\href{http://arxiv.org/abs/1101.0048}{{\tt arXiv:1101.0048 [hep-th]}}].

  \bibitem{Ashoorioon:2014}
A. Ashoorioon and M. M. Sheikh-Jabbari,
  [\href{http://arxiv.org/abs/1405.1685}{{\tt arXiv:1405.1685 [hep-th]}}].

\end{thebibliography}
\end{document}